\documentclass[aps,amsmath,amssymb,12pt]{revtex4}
\usepackage{graphicx,epsf}                           %
\begin{document}
\noindent 

\title{\bf GROUND-STATE CORRELATIONS AND FINAL STATE INTERACTIONS IN THE PROCESS 
\boldmath{ $^3He(e,e'pp)n$}}

\author{C. Ciofi degli Atti}
%
\author{L.P. Kaptari} \thanks{On leave from Bogoliubov Lab. Theor. Phys., JINR, Dubna, Russia }
\affiliation{ 
  Department of Physics, University of Perugia, and
 INFN, Sezione di Perugia,  via A. Pascoli, Perugia, I-06100, Italy}

\date{\today}

\begin{abstract}

  The two-proton emission process $^3He(e,e'pp)n$   is  theoretically investigated using realistic
  three-nucleon wave functions and  taking the  final state interaction into account by  an approach which can be used
   when  the value of the three-nucleon  invariant mass is either below or above the pion emission threshold.
  Various kinematical conditions which enhance  or minimize the effects of  the final state interaction are 
  thoroughly analyzed.

\end{abstract}
\maketitle 
 
\def \bfgr #1{ \mbox {{\boldmath $#1$}}}
\newcommand{\be}{\begin{eqnarray}}
\newcommand{\ee}{\end{eqnarray}}
\newcommand{\nn}{^3He(e,e'N_1N_2)N_3}
\newcommand{\pp}{^3He(e,e'p_1p_2)n}
\newcommand{\pn}{^3He(e,e'p_1n)p_2}
\newcommand{\ga}{\gamma^*}
\newcommand{\emme}{M({\bf p}_{rel}_n,{\bf p}_{rel})}
\newcommand{\spectral}{P_1(k_1,E^*)}
 \topmargin .01cm

\newpage
\section{Introduction}
The investigation of Ground-State Correlations 
(GSC) in nuclei, in particular those which 
originate from the most peculiar features of the 
Nucleon-Nucleon (NN) interaction, i.e. its strong 
short range repulsion and complex state 
dependence (spin, isospin, tensor, etc), is one 
of the most challenging aspects of experimental 
and theoretical nuclear physics and, more 
generally, of hadronic physics. The results of 
sophisticated few- and many-body calculations in 
terms of realistic models of the NN interaction 
(\cite{manybody1,manybody2, pieper}), show 
 that the complex structure of the latter generates a rich correlation structure of the
 nuclear ground-state wave function.
The experimental investigation of the nuclear 
wave function or, better, of various density 
matrices, $\rho (1), \rho(1,1'), \rho(1,2), etc$, 
is therefore necessary in order to 
 ascertain whether the prediction of the {\it Standard Model} of nuclei (structureless non-
 relativistic nucleons interacting via the known NN forces ) is indeed justified in practice, or
 other phenomena, e.g., relativistic effects, many-body forces, medium modification of nucleon properties,
and explicit sub-nucleonic degrees of freedom 
(quark and gluons), have to be advocated in order 
to describe ground-state properties of nuclei at 
normal density and temperature. 

Unfortunately, whereas the one-body density 
matrix (charge density) is experimentally well 
known since many years from elastic electron 
scattering (see e.g.\cite {report}), the present 
knowledge of those quantities which are more 
sensitive to GSC, 
 e.g. the non-diagonal one-body
and two-body density matrices, which could in 
principle be investigated by nucleon ($N$) 
emission processes like, e.g., the $A(e,e'N)X$ 
and $A(e,e'NN)X$ reactions, is still too scarce. 
 The  reasons is  that the  effects from  Final State Interactions (FSI), 
 Meson Exchange Currents (MEC) and Isobar
Configuration (IC) creation, may 
  mask   the
effects generated by GSC. In our view, the 
present situation is such that 
 the longstanding question whether  FSI and other concurrent processes  hinder the investigation of GSC,
  has not yet been
 clearly answered. Moreover, due to the difficulty to treat consistently GSC,  FSI, MEC, {\it etc}  within the full
 complexity of the nuclear  many-body
  approach,
 the answer was in the past merely dictated by philosophical taste  rather than by the results
 of solid calculations and  unambiguous experimental data. 
 A clear cut answer to the above question
 would require, from one side,   realistic many-body calculations of bound and continuum nuclear states, and,
 from the other side, a wise choice of the kinematics and of the  type of  process to be investigated,  
 so as to  possibly minimize all those 
 effects which compete with GSC. In this respect, of particular usefulness 
  is  the two-proton emission process $A(e,e'pp)X$,
   where MEC
 play a minor role (with respect to the proton-neutron emission $A(e,e'pn)X$), since the virtual photon
 does not couple to the exchanged neutral meson, and  IC production is also  suppressed thanks to  
  angular momentum and parity conservation 
 selection rules  (see e.g. \cite{giusti}, \cite{emura}).
  
  The investigation of the two-nucleon
   emission processes has considerably progressed during the last few years, 
  both in the  few-body systems and the
  complex nuclei domains. In the latter case, 
   extensive theoretical studies on the  $A(e,e'pp)X$ process
 have been performed (see e.g. \cite{grab,GP,ryck}   and References therein quoted), aimed at developing  various
 theoretical frameworks to treat GSC and FSI, together with  competing effects, such as  MEC, and, at the same time, 
  experimental 
 data have been obtained  (see e.g.\cite{starink}, \cite{starink1}), which provided non trivial evidence of GSC effects.
  The treatment of the  two-nucleon emission process from few-body systems, which represents  the object of the
 present investigation, has  the non trivial theoretical advantage  that exact 
 ground-state wave functions from variational or Faddeev-type calculations (see e.g. \cite{manybody1,ishi,pisa}
  and References therein quoted) can be used in the  calculations, thus exploiting the whole  realistic
  picture  of GSC; moreover, provided the final three-nucleon invariant mass, $\sqrt {s}$, is below the pion production
  threshold ($\sqrt {s} \simeq 2.95\,\, GeV$), accurate continuum wave functions are also available \cite{GLOCK},
  so that  a fully  consistent 
   treatment of both GSC and FSI effects in the process  $^3He(e,e'pp)n$ at low four-momentum transfer
    has been recently  developed 
   \cite{manybody1,GLOCK} \footnote{From now-on, for 
   {\it low (high) momentum transfer} we will mean $\sqrt {s}$ {\it below (above) 
    the pion threshold}.}.
    Moreover, experimental data at low  momentum transfer ( $Q^2 \sim 0.1\,\, {GeV/c}^2, Q^2 = {\bf q}^2 - \nu^2,
     \nu \sim 0.2 \,\, GeV$)  became available 
   from   NIKHEF \cite{groep}, which made it possible  to produce  a significant comparison between theoretical
    predictions and
    experimental data.
  
  In this paper we are interested in medium and high  momentum transfer regions;
   the reason is twofold:
  $i)$  by increasing the momentum transfer, one might be able to investigate the momentum space wave function
  in a broader kinematical region; $ii)$ processes at high momentum transfer  could provide crucial
  information on the origin and the very mechanism of hadronic rescattering in the medium \cite{tras},
   which has  so far been investigated with simple 
  three-body wave functions. Realistic calculations at intermediate and high values of $Q^2$
  are therefore timely, also in view of  running  experiments at TJLab covering a region of intermediate values of $Q^2$
   ($\nu \sim 0.4 - 1\,\, GeV,\, Q^2 \sim 0.5 -  2\,\, {GeV/c}^2 $)  \cite{CEBAF} .
  It should be reminded, at this point,  that  when the 
  momentum transfer is such that the three-nucleon invariant mass is higher than  the pion production threshold,
   Faddeev-like  calculations in the continuum
  cannot be performed, and the necessity arises of developing a proper treatment of elastic  rescattering
   effects, in presence of inelastic channels.
   It is precisely the aim of this paper to present  such a treatment, 
    and  to thoroughly analyze  the possibility that  
  by a proper choice of the kinematics,  
  the effects of FSI   in the process $^3He(e,e'pp)n$ could  be minimized.
  We would like to stress that our aim is not that  of a direct  comparison with (still lacking) experimental data
  in this region of momentum transfer, since, as previously stated, that would require a proper consideration 
  of effects competing with GSC, but
  rather to try to understand whether  particular kinematical  conditions exist 
  which could minimize the effects from  FSI,  a necessary condition for a meaningful investigation of 
   GSC. 
   Preliminary results of our calculations have already been presented
  in Ref. \cite{preliminary}. 
  Through this paper we shall be using the three-body wave functions obtained by the Pisa Group 
  \cite{pisa,rosati}.

 Our paper is organized as follows: in Section II some general concepts
 concerning the Kinematics
 of the process and the Cross Section will be recalled; our approach to the 
 treatment of FSI is illustrated in Section III,
 together with the results of calculations; the Summary and Conclusions are
 given in Section IV. Some useful formulae concerning two-nucleon correlations in nuclei are given 
 in the Appendix.

 \section{
 Kinematics and cross section}
 We will consider the absorption of a virtual photon ${\gamma^*}$ by  a
 nucleon bound in $^3He$, followed
 by two-nucleon emission, i.e.  the process  $^3He(e,e'N_1N_2)N_3$, where $N_1$ and $N_2$ denote
  the nucleons which
 are detected.  In the rest of this paper
 the photon four momentum transfer will be denoted by 
 \be
  Q^2 =-q^2= -(k_e-k_{e'})^2 = {\bf q}^{\,\,2} - \nu^2=4 {\epsilon}_e  
{\epsilon}_{e'} sin^2 {\theta_e \over 2} 
\label{qu} \ee where $k \equiv (\epsilon,{\bf 
k})$ is the four momentum of the electron, 
 ${\bf q} = {\bf k}_e - {\bf k}_{e'}$, $\nu= {\epsilon}_e - 
{\epsilon_{e^{'}}} $ and $ \theta_e \equiv 
\theta_{\widehat{{\bf k}_e {\bf k}_{e'}}}$. 

 The  momenta of the bound nucleons, before ${\gamma^*}$ absorption, will be denoted  by ${\bf k}_i$,
 and after  ${\gamma^*}$ absorption,  by ${\bf p}_i$. Momentum conservation requires that

 \begin{equation}
 {\sum_{i=1}^3{\bf k}_i} =0 \qquad\qquad  {\sum_{i=1}^3{\bf p}_i = {\bf q}}
 \label{one}
 \end{equation}
and energy conservation that 
\begin{equation}
  \nu +M_3 =\sum_{i=1}^3 (M_N^2+{\bf p}_i^2)^{1/2}
  \label{two}
  \end{equation}
 where $M_N$ and $M_3$ are the masses of the  nucleon and the three-nucleon system, respectively.

\begin{figure}[h] 
\begin{center}
    \includegraphics[height=0.20\textheight]{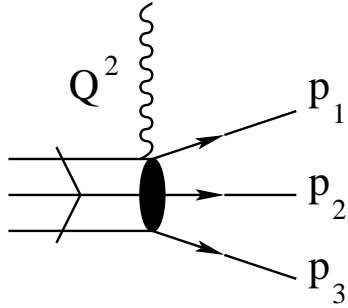}
\caption{The One Photon Exchange diagram for the 
two-nucleon (N) emission off $^3He$, 
    $^3He(e,e'N_1N_2)N_3$. $Q^2$ is the four-momentum transfer and $p_i$ denotes the four- momentum
     of nucleon $N_i$ in the final state. $N_1$ and $N_2$ denote the detected nucleons.}
    \label{fig1}
\end{center}
\end{figure}

 In one-photon exchange approximation,   depicted in Fig. \ref{fig1},
 the cross section of the
process, reads as follows 
\begin{equation}
\frac{d^{12}\sigma} {d\epsilon_{e'} d\Omega_{e'} 
d{\bf p}_1 d{\bf p}_2 d{\bf p}_3 
}={\sigma}_{Mott}\cdot \sum_{\alpha=1}^6 v_\alpha \cdot W_\alpha 
\cdot \delta( {\bf q} - \sum_{i=1}^3{\bf p}_i ) 
\delta( \nu +M_3 -\sum_{i=1}^3 (M_N^2+{\bf 
p}_i^2)^{1/2}) \label{three} 
\end{equation}
where $v_\alpha$ are well known kinematical factors, 
and $W_\alpha$ the {\it response functions}, which 
have the following general form 
\be
W_\alpha =\left | \langle \Psi_f^{(-)}({\bf p}_1, {\bf 
p}_2, {\bf p}_3)|\hat{ \mathcal O}_\alpha({\bf q})| 
\Psi_i({\bf k}_1, {\bf k}_2, {\bf k}_3)\rangle 
\right |^2 \label{four} \ee In Eq.(\ref{four}) 
$\Psi_f^{(-)}({\bf p}_1, {\bf p}_2, {\bf p}_3)$ 
and $\Psi_i({\bf k}_1, {\bf k}_2, {\bf k}_3)$ are 
the continuum and ground-state wave functions of 
the three-body system, respectively, and $\hat{ 
\mathcal O}_\alpha({\bf q})$ is a quantity depending 
on proper combinations of the components of the 
nucleon current operator $\hat{ j^{\mu}}$ (see 
e.g. \cite{report}). Two nucleon emission 
originated by $NN$ correlations can occur because 
 of two
different processes: 
\begin{enumerate}
\item in the initial state $"1"$ and $"2"$ are correlated and $"3"$ is far apart; $\gamma^*$ is
absorbed either by $"1"$ or $"2"$ and all of the 
three-nucleons 
 are emitted in the continuum;
  if nucleon $"3"$  was at rest in the initial state, one has ${\bf k}_1= -{\bf k}_2$  and, if FSI is disregarded,  ${\bf p}_{1(2)}$=${\bf k}_{1(2)}+{\bf q}$, ${\bf p}_{2(1)}$=${\bf k}_{2(1)}$
in the final state; 
\item in the initial state nucleons $"1"$ and   $"2"$ are correlated and $"3"$ is far apart;
$\gamma^*$ is absorbed by $N_3$ and all of the 
three-nucleons are emitted in the continuum. If 
$N_3$ was at rest before interaction, and FSI is 
disregarded, $N_1$ and $N_2$ are emitted 
back-to-back with momenta ${\bf k}_{1}=-{\bf 
k}_{2}$ and ${\bf p}_{3}$=${\bf q}$. 
\end{enumerate}

The above picture is distorted by $FSI$. The aim 
of this paper is precisely to investigate the 
relevance of $FSI$ effects in both processes. 


\section{The  Final State Interaction in the two-nucleon emission process}

We will now assume that $N_{1}$ and $N_{2}$, the 
two detected nucleons, are the two protons 
 and $N_3$ the neutron ($n$). The
two-nucleon emission process will thus be 
${^3H(e,e'p_1p_2)n}$ which, as explained, can 
originate from the two mechanisms described 
above. 

\subsection{Process $1$: absorption of $\gamma^*$ by the correlated $pp$ pair.}

In this process $\ga$ is absorbed by proton "1" 
("2") correlated with proton "2"("1"), and the 
neutron is the "spectator". 

\begin{figure}[h] 
\begin{center}
    \includegraphics[height=0.50\textheight]{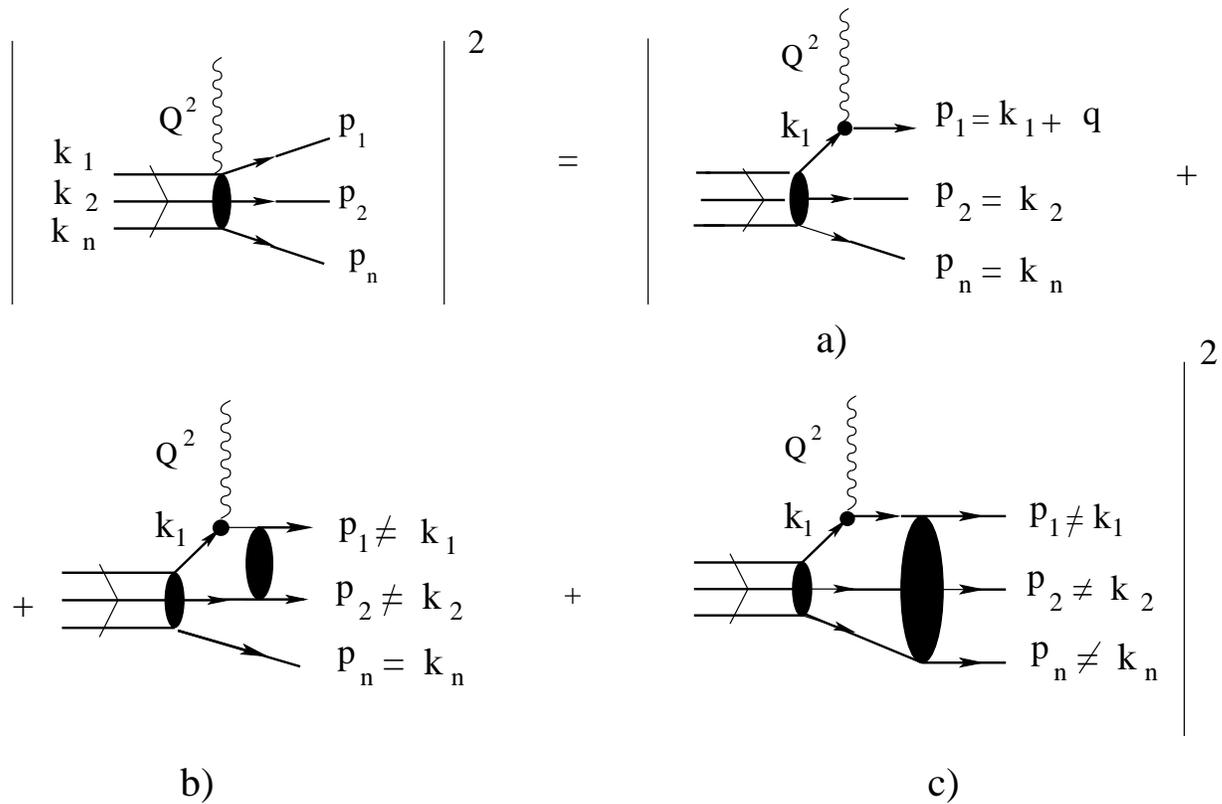}
\vskip -0.5cm \caption{ Schematic representation 
of the various processes contributing to the 
 reaction  $\pp$:
  (a) denotes  the Plane Wave Approximation ({\it PWA}), (b) the {\it pp rescattering},
  (c) the {\it  three-body rescattering}.
   $k_1 (p_1),k_2 (p_2)$  and  $k_n (p_n)$ denote the momenta of proton  $"1"$, proton  $"2"$
  and the neutron, respectively, in the initial (final) state.}
\label{fig2} 
\end{center}
\end{figure}

 The various diagrams, in order of increasing complexity,
 which contribute to the process,  are depicted in
 Fig. \ref{fig2}.

  Let us introduce  the following quantities:
 \begin{enumerate}
 \item the {\em relative momentum} of the detected proton
 pair
 \be
 {\bf p}_{rel}=\frac{{\bf p}_1-{\bf p}_2}{2} \equiv {\bf t}
 \label{six}
 \ee
 \item the {\em Center-of-Mass momentum}  of the pair
 \be
 {\bf P}={\bf p}_1+{\bf p}_2
 \label{seven}
 \ee
\end{enumerate}

In what follows, for ease of presentation, and 
also in order to make the comparison with 
previous calculations more transparent, we will 
consider the effects of the FSI on the 
longitudinal response only. Let us first consider 
diagrams $a)$ and $b)$, i.e. the Plane Wave 
Approximation plus the $pp$ rescattering in the 
final state. By changing the momentum variables 
from 
 ${\bf p}_{1}$ and   ${\bf p}_{2}$  to ${\bf P}$ and  ${\bf p}_{rel}$, and  integrating the cross section  
(Eq. (\ref{three})) over ${\bf P}$ and the 
kinetic energy of the neutron, we obtain 
\be
\frac{d^{8}\sigma}{{d\epsilon}_{e^{'}} 
d{\Omega}_{e^{'}}d\Omega_{p_n} dp_{rel} 
d\Omega_{p_{rel}}} ={\mathcal K}\left( 
Q^2,\nu,{\bf p}_n,{\bf p}_{rel}\right) \cdot 
R_L(\nu, Q^2, {\bf p}_n, {\bf p}_{rel}) 
\label{nine1} \ee with 
\be
R_L(\nu,Q^2, {\bf p}_n, {\bf p}_{rel}) = 
G_E^p(Q^2)\,^2\cdot M^{(pp)}({\bf p}_n, {\bf 
p}_{rel},{\bf q}) \label{rl} \ee where $G_E^p(Q^2)$ 
is the proton electric form factor, 
 ${\mathcal K}$ incorporates all kinematical variables,  and 
 $M^{(pp)}({\bf p}_n,{\bf p}_{rel}, {\bf q})$
is the transition nuclear form factor which 
includes the $pp$ rescattering, {\it viz} 
\begin{eqnarray}
&& M^{(pp)}({\bf p}_n,{\bf p}_{rel},{\bf q})= 
\frac12 \sum\limits_{{\mathcal M}}\, 
\sum\limits_{S_{pp},\Sigma_{pp}}\sum\limits_{\sigma_n} 
|T^{(pp)}({\mathcal 
M},\sigma_n,S_{pp},\Sigma_{pp}, {\bf p}_n, {\bf 
p}_{rel},{\bf q})|^2 \label{oneH} 
\end{eqnarray}
The scattering matrix $T^{(pp)}({\mathcal 
M},\sigma_n,S_{pp},\Sigma_{pp}, {\bf p}_n, {\bf 
p}_{rel},{\bf q} )$ has the following form 

\begin{eqnarray}
&&T^{(pp)}({\mathcal 
M},\sigma_n,S_{pp},\Sigma_{pp}, {\bf p}_n, {\bf 
p}_{rel} ,{\bf q})=\nonumber\\ && \int d^3 r d^3 
\rho \Psi_{{\frac12}{{\mathcal M}}}({\bf 
r},{\bfgr \rho}) \, \chi_{\frac12 
\sigma_n}\exp{(-i{\bf p}_n{\bfgr \rho})} 
  \psi_{S_{pp},\Sigma_{pp}}^{{\bf t}(-)}({\bf r})  
  \exp{(i{\bf qr}/2)},
\label{Ti1} 
\end{eqnarray}

\noindent where $\Psi_{{\frac12}{{\mathcal 
M}}}({\bf r},{\bfgr \rho})$ is the three-nucleon 
wave function, $\psi_{S_{pp},\Sigma_{pp}}^{{\bf 
t}}({\bf r})$ the continuum two-proton wave 
function, and $\chi_{\frac12 \sigma_n}$ the 
neutron spinor. In the above and the following equations, ${\bf 
r}$ and ${\bfgr \rho}$ are the Jacobi coordinates 

\be
{\bf r}= {\bf r}_1 -{\bf r}_2 
\,\,\,\,\,\,\,\,\,\,\,\,\,\,\, {\bfgr {\rho}}= 
{\bf r}_3 - {\frac{1}{2}}({\bf r}_1+{\bf r}_2) 
\label{jac} \ee

When $pp$ rescattering is disregarded, i.e. only 
diagram $a)$ is considered, one has 

\be
{\bf p}_1= {\bf k}_1 +{\bf q} \qquad\qquad {\bf 
p}_2= {\bf k}_2\qquad \qquad {\bf p}_n= {\bf k}_n 
\label{five} \ee 

\be
{\bf p}_{rel}={\bf k}_{rel} +\frac{\bf q}{2} 
\qquad \qquad {\bf P}={\bf K} +{\bf q} 
 \label{five1} 
\ee where 
\be
      {\bf k}_{rel}=\frac{{\bf k}_1 - {\bf k}_2}{2}   \qquad \qquad \qquad    {\bf K}={\bf k}_1 +{\bf k}_2=-{\bf k}_n
      \label{five2}
      \ee
      are the relative and $CM$ momenta of the $pp$ pair  before interaction. 
      The two-proton continuum wave function is simply
  $\psi_{S_{pp},\Sigma_{pp}}^{\bf t}
({\bf r})=\chi_{S_{pp},\Sigma_{pp}} \exp{(i{\bf 
p}_{rel}{\bf r}})$ and the scattering amplitude 
becomes 
\begin{eqnarray}
&&T^{(pp)}({\mathcal 
M},\sigma_n,S_{pp},\Sigma_{pp}, {\bf p}_n, {\bf 
p}_{rel},{\bf q} ) \rightarrow 
T^{(PWA)}({\mathcal 
M},\sigma_n,S_{pp},\Sigma_{pp}, {\bf k}_n, {\bf 
k}_{rel})=\nonumber\\ && \int d^3 r d^3 \rho 
\Psi_{{\frac12}{{\mathcal M}}}({\bf r},{\bfgr 
\rho}) \, \chi_{\frac12 \sigma_n}\exp{(-i{\bf 
k}_n{\bfgr \rho})} 
  \chi_{S_{pp},\Sigma_{pp}} \exp{(-i{\bf k}_{rel} {\bf r})},\nonumber\\
\label{Ti} 
\end{eqnarray}
which is nothing but the three-body wave function 
in momentum space.

The scattering amplitude which include the $pp$ 
rescattering has been calculated using the 
continuum wave function for two interacting 
protons 
\begin{eqnarray}
\psi_{S_{pp},\Sigma_{pp}}^{\bf t}({\bf r})= 
4\pi\,\sum\limits_{lm}\sum\limits_{l'S'J_f} 
\langle\,lm\,S_{pp}\Sigma_{pp}\, |J_fM_J\,\rangle 
i^{l'} {\rm Y}_{lm}^*(\hat{\bf p}_{rel}){\rm 
R}_{lSl'S'}^{J_f}(r){\rm Y}_{l'S'}^{J_fM_J} 
(\hat{\bf r}), \label{wf_cont} 
\end{eqnarray}
where ${\rm Y}_{lm}(\hat{\bf p}_{rel})\left({\rm 
Y}_{l'S'}^{J_fM_J} (\hat{\bf r}) \right) $ 
denotes the spherical (spin-angular) harmonics, 
and ${\rm R}_{lSl'S'}^{J_f}(r)$ is the scattering 
radial wave function, solution of the 
Schr\"odinger equation in the continuum, with 
asymptotic behaviour 
\be
\left. {\rm R}_{lSl'S'}^{J_f}(r)\right 
|_{r\to\infty}\longrightarrow\, 
\delta_{ll'}\delta_{SS'}\exp\left( 
i\delta_l\right)\frac{\sin [tr-(l\pi/2)+\delta_l] 
 }{tr}
\label{asymp} \ee 
 where $t \equiv |{\bf t}|\equiv|{\bf p}_{rel}|$.  In the presence of a  tensor interaction
  the asymptotic  of  ${\rm R}_{lSl'S'}^{J_f}(r)$
is more complicated but, by a unitary 
transformation, other radial wave functions may 
be introduced with asymptotic similar to Eq. 
(\ref{asymp}) (see e.g. Ref. \cite{sitenko}). 

Inserting (\ref{wf_cont}) into (\ref{Ti1}), and 
using the completeness of the scattering wave 
functions, the amplitude $T^{pp}$ in 
Eq.~(\ref{oneH}) can be expressed in the 
following way 
\be
&&T^{(pp)}({\mathcal 
M},\sigma_n,S_{pp},\Sigma_{pp},{\bf p}_n, 
 {\bf p}_{rel},{\bf q} )=
\frac2\pi \int\tilde t^2d\tilde t\, d^3\rho \, 
d^3 r \exp \left(-i{\bf p}_n\bfgr\rho 
\right)\nonumber\\ &&\sum\limits_{\{\alpha\}} 
\langle XM_X\,L_\rho M_\rho|\frac12{\mathcal 
M}\rangle \langle j_{12} M_{12} \frac12 
\sigma_3|XM_X\rangle\nonumber\\ &&\langle l_{12} 
m_{12} S_{pp}\nu |j_{12} M_{12}\rangle \langle 
l_{f} m_{f} S_{pp}\Sigma_{pp}|j_{f} M_{f}\rangle 
\langle l_{f} \tilde m_{f} S_{pp}\nu |j_{f} M_{f} 
\rangle \nonumber \\ &&{\rm Y}_{l_{f} m_{f}}(\hat 
{\bf p}_{rel}) {\rm Y}_{l_{12}m_{12}}(\hat {\bf 
r}) {\rm Y}_{l_{f}\tilde m_{f}}^* (\hat {\bf 
r})\nonumber\\ &&\exp (i{\bf qr}/2){\rm 
R}_{l_{12}S_{pp}}^{j_{12},\tilde t}(r) {\rm 
R}_{l_{f}S_{pp}}^{j_{f}, t}(r) 
(-i)^{l_f}I^{|\tilde {\bf p}_{rel}|}_{\{\alpha 
\}}(|\bfgr\rho|), \ee where ${\{\alpha \}}$ 
denotes the full set of quantum numbers 
characterizing the ground-state partial 
configurations in the $^3He$ wave function, and $ 
I^{|\tilde {\bf p}_{rel}|}_{\{\alpha 
\}}(|\bfgr\rho|)$ are the corresponding overlaps 
with the scattering state wave functions. 

\begin{figure}[h] 
\begin{center}

    \includegraphics[height=0.5\textheight]{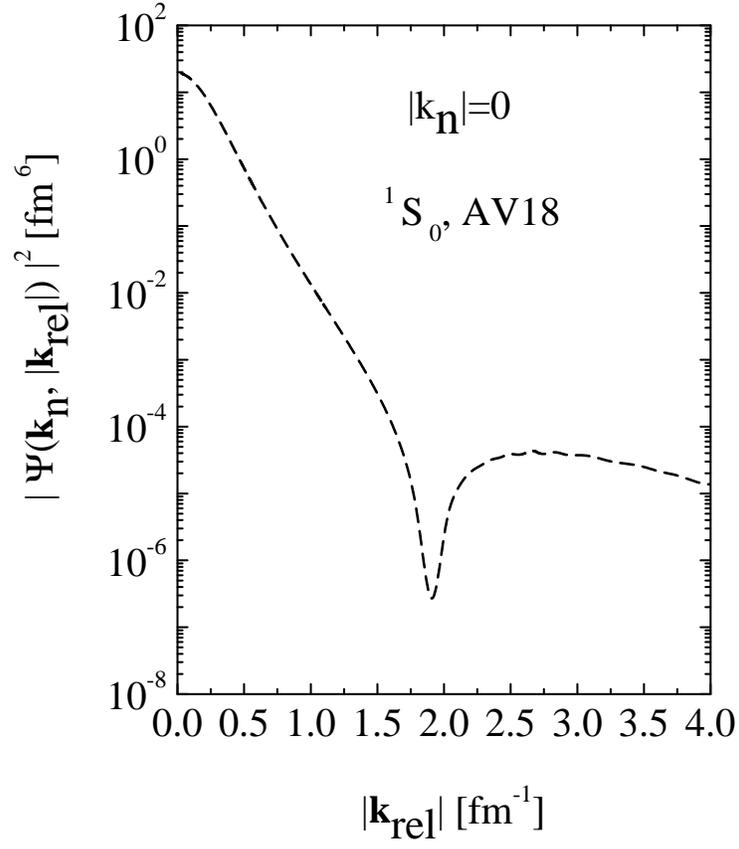}
\caption{ The momentum space wave function of 
$^3He$ corresponding to the configuration in 
which the neutron is at rest and the two protons 
are in the state $^1S_0$ of relative motion with 
momentum ${\bf k}_{rel}=({\bf k}_1-{\bf k}_2)/2$. 
Three-body wave function from Ref. \cite{rosati}; 
$AV18$ interaction \cite{AV14}.} \label{fig3} 
\end{center}
\end{figure}

In what follows the 
 so-called {\it symmetric (sym)} kinematics \cite{Laget}
\begin{eqnarray*}
{\bf p}_n=0; \quad 
  {\bf p}_1 + {\bf p}_2= {\bf q}
  \end{eqnarray*}
  will be considered, which  corresponds, in $PWA$, to a ground-state configuration 
  characterized by the two protons with  equal and opposite momenta and
  the neutron with  zero momentum.
  In the {\it sym} kinematics, the transition form factor
  $M^{(pp)}({\bf p}_n,{\bf p}_{rel},{\bf q})$, when $\gamma^*$ interacts with proton $"1"("2")$, will only depend upon
  ${\bf p}_{rel}={\bf q}/2 - {\bf p}_{2(1)}$,  i.e., for a fixed
  value of $|{\bf q}|$, will only depend  upon 
    $|{\bf p}_{2(1)}|$ and the angle between ${\bf q}$ and  ${\bf p}_{2(1)}$. In   PWA,  
 when the angular momentum
  of the neutron is zero, also the $pp$ pair has  relative angular momentum zero, so that the   cross section is
  almost entirely determined
  by the square of the $^1S_0$ component of the three-body wave function
  $\Psi({\bf k}_{rel},{\bf k}_{n}=0)$,  which is   shown in
  Fig.\,\ref{fig3}.
  
  The calculated transition form factor   is
  shown  in Fig.\ref{fig4}. Calculations have been performed with  three-body  wave function obtained in
ref. \cite{rosati} using the AV18 interaction 
\cite{AV14}. Our results, which are in agreement 
with the ones of Ref. \cite{Laget} (where a 
different ground-state wave functions has been 
used), show that the $pp$ rescattering is very 
large and completely distorts the $PWA$ results. 

 In order to investigate to what extent $FSI$ depend 
 upon the kinematics of the process,
we have also considered the {\it super-parallel 
(s.p.)} kinematics, according to which one still 
has 
 ${\bf p}_n=0$,
  ${\bf p}_1 + {\bf p}_2= {\bf q}$,  but  all momenta are collinear, i.e.( ${\bf q} \parallel z$)
  \be
  {\bf p}_{1\perp}={\bf p}_{2\perp}=0 \qquad\qquad\qquad p_{1z}+p_{2z}=|\bf q|
  \ee

The results of calculations, which are presented 
in Fig.\ref{fig5},\, look very different from the 
ones shown in Fig. \ref{fig4}. Concerning these 
differences, the following remarks are in order: 
 \begin{enumerate}
  \item as far as the $PWA$ results are concerned, it can be seen that the transition matrix elements  differ, 
  at the same value of $\nu$ ,
 by more than one order of magnitude; the reason is that the relative
  momentum $|{\bf k}_{rel}|$, at a given value of $\nu$,  is very different  in  the two kinematics, with
  $|{\bf k}_{rel}^{(sym)}| \gg  |{\bf k}_{rel}^{(s.p.)} |$
  (e.g,  at $\nu$ = $0.2\, GeV$ one has $|{\bf k}_{rel}^{(s.p.)}|\simeq 0.45\ fm^{-1}$,  whereas
  $|{\bf k}_{rel}^{(sym)}|\simeq 2.2\ fm^{-1}$). Since in both kinematics $M^{PWA}({\bf k}_n,{\bf k}_{rel})$ 
  is entirely determined by the
  $^1S_0$ three-body
  wave function $\Psi(|{\bf k}_{rel}|,{\bf k}_n=0,)$,  the value of $\nu$ at which
  this exhibits its  minimum is different
  in the two cases;
 
  \item
   in Fig. \ref{fig4} the $pp$ rescattering  effects depend upon the value of  the relative
   momentum  of the two  protons in the final state 
    $|{\bf p}_{rel}|$ = $|{\bf p}_1|\sin\displaystyle\frac{\theta_{12}}{2} \neq |{\bf k}_{rel}|$.
   Unlike the $PWA$ case, one has $|{\bf p}_{rel}^{({\it sym})}| \leq |{\bf p}_{rel}^ {({\it s.p})}|$
   (e.g.  at $\nu = 0.2\, GeV$ one has  $|{\bf q}| \simeq 535 MeV/c$,
   $|{\bf p}_2| \simeq 0.45\ fm^{-1} $ and  $|{\bf p}_{rel}|\simeq 1.8\ fm^{-1}$,
   in  the {\it s.p.}  kinematics,  and     
   $|{\bf p}_1|=|{\bf p}_2| \simeq 2.2\ fm^{-1} $, $\theta_{12}\simeq 105^0$
   and $|{\bf p}_{rel}|=|{\bf p}_1|\sin\displaystyle\frac{\theta_{12}}{2}\simeq 1.7\ fm^{-1}$
   in the {\it sym} kinematics). Thus, the two-proton relative energy in the final state is larger
   in the {\it s.p.} kinematics, which explains the apparent smaller effects of $FSI$ in Fig. \ref{fig5}.
   In this respect, it should however be pointed out that at values
    of $\nu > 0.7-0.8\ GeV$, i.e. at large values of  $p_{2z} \geq 1 fm^{-1}$, where
    correlation effects are more relevant, the momentum  transfer $|{\bf q}|$  and 
   the relative momentum of the proton pair  become
   very large, and the Schr\"odinger equation can not in principle be applied 
    to describe the $pp$-interaction in the continuum (e.g. in the {\it s.p.} kinematics, when 
   $|{\bf q}|\geq 1\ GeV/c$, $|{\bf p}_{2}|\simeq 0.5 \,GeV/c $, $|{\bf p}_{1}|\simeq 1.5\, GeV/c$,
   $|{\bf p}_{rel}|\simeq 1 GeV/c $). To treat the case of high energies,  a  Glauber-type calculation is in progress and
   will be reported elsewhere \cite{ckm}. Thus, it appears that in the {\it s.p.}   kinematics
   considered in  Fig. \ref{fig5},  there exists  only a small bin of $\nu \simeq 0.4-0.5 \,GeV$
    where two-nucleon correlations could be investigated treating the $pp$ rescattering within
    the Schr\"odinger equation;
    \end{enumerate}

The next contribution to be considered is the 
proton-neutron rescattering (diagram (c) in Fig. 
\ref{fig2}). This has been found in Ref. 
\cite{Laget} to provide very small effects, as 
also recently found in Ref. \cite{preliminary}. 
This point will be discussed in details in the 
next Section. 

\subsection{Process 2: absorption of $\gamma^*$ by the neutron.}
\subsubsection{The Plane Wave Approximation and the $pp$ rescattering}

We will now consider the process $\pp$ , in which 
$\ga$ 
 interacts with the neutron and the two protons are  emitted and detected. We will consider
  two extreme cases of this process, {\it viz} : i) in the initial state the neutron is a partner of a correlated
 proton-neutron pair, with the second proton far apart from the pair;
  ii) in the initial state the neutron is at rest, far apart from the
 two correlated protons.
 Process ii) has been considered in Ref. \cite{GLOCK} for the case of both a neutron and a proton at rest in the
 initial state. We will  compare  our results with the ones  of Ref. \cite{GLOCK}, considering  only the case
 of the  neutron at rest.  The various mechanisms, in order of increasing complexity,
 which contribute to the above process,  are depicted in
 Fig. \ref{fig6}.

 When  the final state rescattering
 between the two protons is taken into account  (processes  $a)$ plus process  $b)$), but 
 the interaction of the hit neutron with the emitted proton-proton
  pair is disregarded, one has ${\bf p}_{n}={\bf k}_{n}+ {\bf q}$, and 
 the  cross section (Eq. (\ref{three})) integrated 
 over ${\bf P}$ and the kinetic energy of the neutron
  exhibits the same structure of Eq. (\ref{nine1}),
 with $R_L$
given by 

\be
 R_L(\nu, Q^2, {\bf p}_n, {\bf p}_{rel}) = G_E^n(Q^2)\,^2\cdot M^{(pp)}({\bf p}_n, {\bf p}_{rel},{\bf q})
\label{rl1} \ee 

\noindent which differs from Eq. (\ref{rl}) in two 
respects: i) the proton electric form factor is 
replaced by the 
 neutron  one $G_E^n(Q^2)$; ii)
 $M^{(pp)}({\bf p}_n,{\bf p}_{rel}, {\bf q})$
  includes  
the rescattering {\it between the two spectator 
protons} and not between the active and recoiling 
protons; this means that 
   $T^{pp}$
has the following form 
\begin{eqnarray}
T^{pp}({\mathcal M},\sigma_n,S_{pp},\Sigma_{pp}, 
{\bf k}_n, {\bf p}_{rel} )= \int 
 d^3 \rho \exp{(-i{\bf k}_n{\bfgr \rho})}\chi_{\frac12 \sigma_n}
  I_{ {\mathcal M},S_{pp},\Sigma_{pp}}^ {{\bf t},pp}({\bfgr \rho})
\label{new2} 
\end{eqnarray}
 
\noindent where ${\bf k}_n = {\bf p}_n-{\bf q} = 
-({\bf p}_1+{\bf p}_2)$, and 
  $I^{\bf t}$ is the {\it overlap  integral}
between the three-nucleon ground-state wave 
function and the two-proton 
 continuum state,
 i.e.
\be
&& 
 I_{ {\mathcal M},S_{pp},\Sigma_{pp}} ^{{\bf t},pp}({\bfgr \rho})=\int \psi_{S_{pp},
 \Sigma_{pp}}^{{\bf t}(-)}({\bf r})\Psi_{{\frac12}{{\mathcal M}}}
 ({\bf r},{\bfgr \rho})\,
d^3 r \label{eleven} \ee

\noindent We reiterate that in the process analyzed in the 
previous Section, the $pp$ rescattering occurred 
between the protons of the {\it active pair} 
which absorbed the virtual photon, whereas in 
this case $\ga$ is absorbed by the neutron and 
the $pp$ rescattering involves the two {\it 
spectator} protons. 
Within the PWA, \,i.e. when only process $a)$ 
contributes to the reaction, one has 

$\psi_{S_{pp},\Sigma_{pp}}^{{\bf t}} ({\bf 
r})=\chi_{S_{pp},\Sigma_{pp}} \exp{(i{\bf t}{\bf r})}$ 

so that 

\be
 I_{N_2N_3}^{{\bf t},PWA}({\bfgr \rho})=\int \Psi_{{\frac12}{{\mathcal M}}}({\bf r},
 {\bfgr \rho})\,
 \chi_{S_{pp},\Sigma_{pp}} \exp{(-i{\bf t}{\bf r}}) d^3 r
\label{eleven1} \ee 

\noindent and 
\begin{eqnarray}
&&T^{PWA}({\mathcal 
M},\sigma_n,S_{pp},\Sigma_{pp}, {\bf k}_n, {\bf 
k}_{rel} )=\nonumber\\ && \int d^3 r d^3 \rho 
\Psi_{{\frac12}{{\mathcal M}}}({\bf r},{\bfgr 
\rho}) \, \chi_{\frac12 \sigma_n}\exp{(-i{\bf 
k}_n{\bfgr \rho})} 
  \chi_{S_{pp},\Sigma_{pp}} \exp{(-i{{\bf k}_{rel}} {\bf r})},
\label{new3} 
\end{eqnarray}
 which, as in $Process 1)$,  is nothing but 
   the three-nucleon wave function in momentum space.
  
  It is interesting to point out that 
the integral of the transition form factor 
(\ref{new2}) over the 
 direction of ${\bf p}_{rel}$,
 is related to the neutron {\it Spectral Function} $P_1(k_n,E^*)$ \cite{CPS}
 \be
 && P_1(k_n,E^*)=\nonumber\\
 &&\frac{|{\bf p}_{rel}|M_N}{(2\pi)^4}
 \sum\limits_{{\mathcal M},S_{pp},\Sigma_{pp},\sigma_n} \int d \Omega_t\left| \int
d^3r d^3 \rho \exp{(-i{\bf k}_n{\bfgr 
\rho})}\chi_{\frac12 
\sigma_n}\psi_{S_{pp}\Sigma_{pp}}^{{\bf 
t}(-)}({\bf r}) 
  \Psi_{{\frac12}{{\mathcal M}}}
 ({\bf r},{\bfgr \rho})
\right|^2 
 \label{P1}
 \ee
 by the following relation
 
 \be
\int{\mathcal P}_1({\bf k}_n, {\bf p}_{rel}) 
d\Omega_{{\bf p}_{rel}}=P_1(k_n,E^*) 
\label{5teen} \ee where 
\be
{\mathcal P}_1({\bf k}_n, {\bf p}_{rel})= 
\frac{M_N|{\bf p}_{rel}|}{2} M^{pp}({\bf k}_n, 
{\bf p}_{rel}) 
  \label{5teen1}
\ee 
 will be called here the {\it Vector Spectral Function},  and
\be E^*=\frac{{\bf p}_{rel}^2}{M} = \frac{({\bf 
p}_1 -{\bf p}_2)^2}{4M} \label{5teen2} \ee 
 is the "excitation energy" of the spectator $pp$ pair , which is related to the 
 {\it neutron  removal energy} $E$
by the relation 
 $
 E= E_3 + E^* $,
 where $E_3$ is the (positive) binding energy of the three-nucleon system
 (cf. Ref. \cite{CPS}).
 
 The cross section for the two-proton emission process  can then be written
 in the following form
 \be
\frac{d^{8}\sigma}{d{\epsilon}_{e^{'}} 
d{\Omega}_{e^{'}} d\Omega_{p_n} d{p_{rel}} 
 d\Omega_{p_{rel}}} = {\mathcal K}\left( Q^2,\nu,{\bf p}_n,{\bf p}_{rel}\right)
\, \cdot 2\frac{G_E^n(Q^2)}{M_N|{\bf 
p}_{rel}|}{\mathcal P}_1({\bf k}_n, {\bf 
p}_{rel}) \label{crossvec} \ee By integrating 
over $\Omega_{p_{rel}}$, 
 one obtains  the cross section for the semi-inclusive process   
$^3He(e,e'n)pp$ 
\be
\frac{d^{6}\sigma}{d{\epsilon}_{e^{'}} 
d{\Omega}_{e^{'}} d\Omega_{n} d{p_{rel}}} = 
{\mathcal K}G_E^n(Q^2)^2\cdot P_1(k_n,E^*) 
\label{8teen} \ee

 The neutron 
 Spectral Function calculated with and without the $pp$ rescattering \cite{CPS}
 is shown in Fig. \ref{fig7}.
  It can be seen that at    $ k_n \geq 1.5 fm^{-1}$   there are
  regions,  peaked at $E^* \simeq 
  \displaystyle\frac{{{\bf k}_n}^2}{4M}$,   where the $pp$  rescattering does not play any role;
   since  $E^* = \displaystyle\frac{{{\bf p}_{rel}}^2}{M}$, at the peaks 
we have the following relation between $|{\bf 
k}_n|$ and $|{\bf p}_{rel}|$ 

\be |{\bf k}_n| \simeq 2|{\bf p}_{rel}| 
\label{9teen} \ee 

 The positions of the  peaks have been
originally interpreted as arising from a {\it 
two-nucleon correlation} $(2NC)$ configuration, 
where proton $"1(2)"$ is correlated with the 
neutron with momenta satisfying the following 
relations: ${\bf k}_{2(1)}=0$, ${\bf k}_{1(2)} 
\simeq -{\bf k}_n$ \cite {FS}; moreover, 
 the energy
dependence around the peaks can be related to the 
motion of the third particle ${\bf k}_{2(1)} \neq 
0$, or, equivalently, to the Center-of-Mass 
motion of the correlated pair \cite{2NC}. 
 The existence  of the region where $pp$ rescattering is vanishing,  is a general feature of any
 Spectral Function, independently of the two-nucleon interaction and of
 the method to generate the
 wave function.
 This is illustrated in Fig. \ref{fig4},  where the Spectral Function  corresponding
 to the variational three-body  wave function of Ref. \cite{rosati}  obtained  with 
  the $AV18$ 
 interaction \cite{AV14}, is shown for
 several  values of $k \equiv |{\bf k}_n|$.
 It can be  seen that for values of $k_n$ which satisfy  relation (\ref{9teen}),
    FSI effects
 generated by $pp$ rescattering are negligible.
 
 Relation (\ref{9teen}) does not necessarily imply a $2NC$ configuration,
  i.e.
 ${\bf k}_{2(1)}=0$,  ${\bf k}_{1(2)} \simeq -{\bf k}_n$, for it holds also if
  ${\bf k}_{2(1)} \neq 0$,  ${\bf k}_{1(2)} \neq -{\bf k}_n$. The analysis of 
   the {\it Vector Spectral Function} ${\mathcal P}_1(|{\bf k}_n|, |{\bf p}_{rel}|, \theta)$
   should tell us what are the dominant configurations in the three-body wave function.
    To this end,  we have plotted the  {\it Vector Spectral Function} {\it vs} $\theta$
 in 
correspondence of two values of $|{\bf 
p}_{rel}|$: $|{\bf p}_{rel}|=0.75 fm^{-1}$ ( 
$E^{*}=|{\bf p}_{rel}|^2/M \simeq 23 MeV$) and 
$|{\bf p}_{rel}|=1.1 fm^{-1}$ 
 ($E^{*}=|{\bf p}_{rel}|^2/M \simeq 50 MeV$)
  and  various values of $k_n$. The results, which are shown in
  Fig.~\ref{fig9}, deserve the following comments:
\begin{enumerate}
\item  
the constant behaviour of ${\mathcal P}_1$ at 
$k_n$ =0 can easily be understood 
 by considering that, as previously
discussed, only the angle independent $^1S_0$ 
$pp$ 
 wave function contributes;
 \item   
 Eq. (\ref{9teen}), i.e. the relation $|{\bf k}_n|=2|{\bf p}_{rel}|$, would
 correspond to $|{\bf k}_n| \simeq 1.5,  fm^{-1}$, when  $|{\bf p}_{rel}|=0.75 fm^{-1}$, and to
  $|{\bf k}_n| \simeq 2.2 fm^{-1}$, when   $|{\bf p}_{rel}|=1.1 fm^{-1}$. It can be seen that 
 when   $|{\bf k}_n|=2|{\bf p}_{rel}|$  $pp$ rescattering  effects almost disappear, but they become extremely large
 when such a relation is not satisfied, except when  
   $|{\bf k}_n| \geq 2|{\bf p}_{rel}|$  {\it and} $\theta \simeq 0^{o}, 180^{o}$;
 \item 
 when 
  $|{\bf k}_n|=2|{\bf p}_{rel}|$, the value  $\theta =0^{0}(180^{0})$ ({\it super-parallel kinematics}),
  corresponds to  $|{\bf k}_{1(2)}|$=0, ${\bf k}_n=-{\bf k}_{2(1)}$,
 i.e. to the $2NC$ configuration (see Appendix). The results presented in fig. \ref{fig9} clearly
 show that such a configuration is the dominant one; as a matter of fact, far from such
 a configuration (e.g. at  $\theta=90^{0}$, when three-nucleon configurations are important), 
 the {\it Vector Spectral Function} 
is sensibly smaller. 
  Thus,  the most probable configuration in the three-nucleon wave function, when   $|{\bf k}_n|=2|{\bf p}_{rel}|$,
   is indeed  the $2NC$
  configuration,  when one nucleon of the $pp$  pair is  almost at rest and the second one has momentum almost equal
  and opposite to the momentum of the neutron;
  \item
   when   $|{\bf k}_n|\neq 2|{\bf p}_{rel}|$ and $\theta \simeq 0^{o},180^{o}$, 
     we still stay in the {\it super-parallel kinematics} but not in the 
  {\it two-nucleon correlation} region, for now  $|{\bf k}_{2(1)}| \neq 0$, ${\bf k}_n \neq -{\bf k}_{1(2)}$; however,
  it can be seen that for   $|{\bf k}_n|>2|{\bf p}_{rel}|$ ,  when the violation 
  of the condition
    $|{\bf k}_n|=2|{\bf p}_{rel}|$ is very mild, $pp$ rescattering effects are still very small; note, moreover,
    that if the $FSI$ can only be described by the $pp$ rescattering, the cross section should be the same  at both angles, 
    a behaviour which deserves experimental
    investigation.
  \end{enumerate}

  Thus the study of the  {\it Vector
 Spectral Function} at   $|{\bf k}_n|\simeq 2|{\bf p}_{rel}|$, $\theta \simeq 0^{o}$ or $180^{o}$, which 
 could be undertaken by  measuring the 
 $(e,e'2p)$ process in the {\it super-parallel  kinematics}, would allow one 
 to obtain information on the three-nucleon wave function in momentum space, provided  the rescattering
 of the two protons with the outgoing neutron does not appreciably distort the process.
  The full calculation of the transition matrix element at low momentum transfer,  has been undertaken 
  in Ref. \cite{GLOCK} within a consistent
  Faddeev approach to bound and continuum states of the three-nucleon system. The process considered
   in Ref. \cite{GLOCK} 
  is the absorption of $\ga$ by a neutron (proton) at rest with the two protons (proton-neutron) emitted back-to-back with
   equal
  momenta ${\bf p}_1 =- {\bf p}_2 \equiv {\bf p}$ and the neutron with momentum
   ${\bf p}_n ={\bf q}$ emitted in a direction perpendicular to ${\bf p}$.
     Thanks to the fully consistent
    treatment of bound
    and continuum state wave functions, the  calculation presented in Ref. \cite{GLOCK},
    represents  the status-of-the-art of the description  of   $Process\, 2)$ 
   at low momentum transfer. 
   In order to extend the theoretical description of the two-nucleon emission processes to the
    high momentum transfer region, where three-body  continuum Faddeev-like wave functions are not yet available,
 we have   developed an approach to  the three-body rescattering, to be presented in  the next subsection,  based
 upon the eikonal approximation, which not only allows one to calculate the high momentum transfer processes, but can  
 also  easily be extended to complex nuclei.

 \subsection{The three-body rescattering}

We have considered the three-body rescattering of 
the neutron with the interacting $pp$ pair 
 within an extended Glauber-type
approach \cite{glauber} based on the following 
assumption 
\be
\chi_{\frac12 \,\sigma_n} \exp(-i{\bf p}_n{\bf 
r}_3)\, \Psi_{S_{pp}\Sigma_{pp}}^{{\bf 
p}_{rel}}({\bf r}_1,{\bf r}_2)\longrightarrow 
\hat G({\bf r}_1,{\bf r}_2,{\bf r}_3) 
\,\chi_{\frac12 \,\sigma_n} \exp(-i{\bf p}_n{\bf 
r}_3)\, \Psi_{S_{pp}\Sigma_{pp}}^{{\bf 
p}_{rel}}({\bf r}_1,{\bf r}_2) \label{glab1} \ee 
where the Glauber operator $\hat G$ is 
\cite{nikolaev} 
\be
\hat G({\bf r}_1,{\bf r}_2,{\bf 
r}_3)=\prod_{i=1}^2\left[ 1-\theta(z_i-z_3)\Gamma 
({\bf b}_i-{\bf b}_3)\right], \ee with ${\bf 
b}_i$ and $z_i$ being the transverse and the 
longitudinal co-ordinates of the nucleon, and 
$\Gamma({\bf b})$ the profile function of the 
elastic nucleon-nucleon scattering amplitude; for 
the latter, the standard high-energy 
parametrization {\it viz}, 
\be
\Gamma({\bf b})= 
\frac{\sigma_{NN}^{tot}(1-i\alpha_{NN})}{4\pi 
b_0^2} 
{\exp{\left(-\displaystyle\frac{b^2}{2b_0^2}\right)}} 
\label{glaub2} \ee has been used, where 
$\sigma_{NN}^{tot}$ is the total Nucleon-Nucleon 
($NN$) cross section and $\alpha_{NN}$ the ratio 
of the imaginary to the real part of the $NN$ 
scattering amplitude. Within such an approach, 
the full distorted transition matrix element 
assumes the following form 
\be
M^{D}({\bf p}_m,{\bf p}_{rel})= \frac12 
\sum\limits_{{\mathcal M}}\, 
\sum\limits_{S_{pp},\Sigma_{pp}}\sum\limits_{\sigma_n} 
|T^{D}({\mathcal M},\sigma_n,S_{pp},\Sigma_{pp}, 
{\bf p}_m, {\bf p}_{rel})|^2 \label{MG} \ee where 
\be
{\bf p}_{m}={\bf p}_{n}-{\bf q}=-({\bf 
p}_{1}+{\bf p}_{2}) \label{pm} \ee is the {\it 
missing momentum}, which coincides with the 
neutron momentum before interaction when the 
three-body rescattering is disregarded. The 
distorted scattering matrix $T^{D}$ has the form 
\be
T^{D}({\mathcal M},\sigma_n,S_{pp},\Sigma_{pp}, 
{\bf p}_{n},{\bf p}_{rel})= \frac2\pi \int\tilde 
t^2d\tilde t\, d^3\rho \, d^3 r \exp \left(-i{\bf 
p}_m\bfgr\rho \right) \Omega^{\tilde t,{\bf 
p}_{rel}}_{{\mathcal 
M},\sigma_n,S_{pp},\Sigma_{pp}}({{\bf r},\bfgr 
\rho}), \label{glaub3} \ee with 
\be
&&\Omega^{\tilde t,{\bf t}}_{{\mathcal 
M},\sigma_n,S_{pp},\Sigma_{pp}}({{\bf r}, \bfgr 
\rho})=\nonumber\\ &&\sum\limits_{\{\alpha\}} 
\langle XM_X\,L_\rho M_\rho|\frac12{\mathcal 
M}\rangle \langle j_{12} M_{12} \frac12 
\sigma_n|XM_X\rangle \langle l_{f} m_{f} 
S_{pp}\Sigma_{pp}|j_{f} M_{f}\rangle \nonumber \\ 
&& {\rm Y}_{l_{f} m_{f}}(\hat {\bf p}_{rel}) {\rm 
Y}_{L_{\rho} M_{\rho}} (\hat {\bfgr \rho}) 
I^{|\tilde {\bf t}|}_{\{\alpha \}}(|\bfgr\rho|) 
\left[ {\rm R}_{l_{f}S_{pp}}^{j_{f},\ t}(r) {\rm 
R}_{l_{12}S_{pp}}^{j_{12},\ \tilde t}(r) \right] 
\left [ {\rm Y}_{l_{12}S}^{j_{12}M_{12}}(\hat 
{\bf r}) {\rm Y}_{l_{f}S_f}^{*\, j_{f}M_{f}}(\hat 
{\bf r}) \right] \hat G({\bf r},\bfgr \rho), 
\label{glaub4} \ee 

\noindent and the response is given by Eq. 
\ref{rl1}, with $M^{pp}({\bf p}_n,{\bf p}_{rel})$ 
replaced by $M^{D}({\bf p}_n,{\bf p}_{rel})$ 
  \footnote{The quantity  ${\mathcal P}_1^D({\bf k}_n, {\bf p}_{rel})$=
  ${M_N|{\bf p}_{rel}|} M^{D}({\bf k}_n, {\bf p}_{rel})/2$ can be called the
   {\it Distorted Vector Spectral Function}}.

 Some details of our numerical calculations of 
 the three-body rescattering transition form factor, Eq.\ref{glaub3},
  are now in order. It can be seen that  the dependence of
  $T^{D}({\mathcal M},\sigma_n,S_{pp},\Sigma_{pp}, {\bf p}_{n},{\bf p}_{rel})$ upon 
  ${\bf p}_{rel}$ is entirely governed 
  by the quantity 
  $\Omega^{\tilde t,{\bf t}}_{{\mathcal M},\sigma_n,S_{pp},\Sigma_{pp}}({{\bf r}, \bfgr \rho})$.
   Since 
  the main component of the $^3He$ ground-state wave function corresponds to the relative motion of the
  $pp$ pair in the   $^1S_o$ wave, the two protons in the final state  are
   mostly in states with spin $S=0$ and even values of the relative angular momenta; thus 
 $\Omega^{\tilde t,{\bf t}}_{{\mathcal M},\sigma_n,S_{pp},\Sigma_{pp}}({{\bf r}, \bfgr \rho})$
 is almost symmetric under the exchange ${\bf t}\leftrightarrow -{\bf t}$. This  symmetry 
 can be  slightly
 violated due to the contribution of the
  highest partial $pp$-waves in  the $^3He$ wave function \cite{nikolaev}. 
  
  The quantities  $\sigma_{NN}^{NN}$, $\alpha_{NN}$ and $b_0$ in Eq. (\ref{glaub2})  
 depend in principle on the total energy of the interacting nucleons \cite{glauber}, however
 at high values of $|{\bf p}_n|$ ($|{\bf p}_n| \ge 0.7 GeV/c$, which implies high values of the 
 momentum transfer $|{\bf q}|$), they become energy independent and
 the "asymptotic" values  $\sigma_{tot}(NN)\sim 44\ mb$, $\alpha_{NN}\simeq -0.4$ can be used,  with  
  $b_0$ determined by $\sigma_{NN}^{NN}$ and $\alpha_{NN}$ from unitarity requirements
 \cite{glauber}.

 The transition form factor  $M^D({\bf p}_n, {\bf p}_{rel})$
    is shown  Figs.\ref{fig10}, \ref{fig11}  and \ref{fig12}, 
 where it is compared with $M^{PWA}$ and $M^{pp}$. 
  The three Figures correspond to three different
 kinematical conditions, namely:
 \begin{enumerate}
 
 \item  Fig. \ref{fig10}  shows the results obtained in the {\it super-parallel kinematics}
    ($\theta_1
=180^{o}$) and $|{\bf p}_{rel}|=0.75 fm^{-1}$, 
{\it vs} $|{\bf P}|= |{\bf p}_{1}+{\bf p}_{2}|= 
 |{\bf p}_{n}-{\bf q}|$.
  Let us, first of all,  discuss the results obtained within the   
  $PWA$ and the $PWA$ plus $pp$ 
 rescattering 
  (dashed
 and dot-dashed curves, respectively), which obviously coincide with
 the results presented   in Fig. \ref{fig9} (top panel, $\theta_1 =180^{o}$), since 
   ${\bf p}_{n}= {\bf k}_{n}+{\bf q}$ and
 ${\bf P}={\bf p}_{mis}=-{\bf k}_{n}$.
The arrow 
 denotes the $2NC$  kinematics, when  ${\bf k}_1= -{\bf k}_{n}$, ${\bf k}_{2}
 =0$ (${\bf p}_{1}+{\bf p}_{n}$=${\bf q}$, ${\bf p}_{2}=0$)
 and $|{\bf k}_{n}| = 2|{\bf p}_{rel}|$.
 In agreement with
 Fig. \ref{fig9}, we see  that the $pp$ rescattering has large effects  at low values of $|{\bf P}|=
 |{\bf k}_{n}|$, but gives negligible contributions  when $|{\bf P}|=|{\bf k}_n| \geq 2|{\bf p}_{rel}|=1.5 fm^{-1}$. 
The full line in Fig. \ref{fig10} includes the 
effects from 
 $n-pp$ rescattering (when ${\bf p}_{m}\neq
 -{\bf k}_{n}$); these effects are  very large at small values of $|{\bf P}|$, but  becomes negligible at
 $|{\bf P}| \geq 2|{\bf p}_{rel}|$. One could be tempted to compare the results shown in Fig. \ref{fig10} with the
 ones presented in Fig. \ref{fig5}. In this respect one should first of all stress that in the {\it sym} kinematics
 used in Fig. \ref{fig5}, the {\it neutron}
 is at rest  (both in the initial and final states),  which means that the transition matrix element is mainly
 governed by the $^1S_0$ wave function of the two-proton relative motion. In the process considered in 
 Fig. \ref{fig10}, when  $\ga$ couples to the neutron, none of the nucleons are at rest, except nucleon "2"
 in the particular kinematics denoted by the arrow which represents the absorption of $\ga$ by a neutron of a correlated $np$ pair, with the spectator {\it proton} at rest both in the initial
 and the final states.  
 In this case, the transition matrix element gets contributions from
   higher angular momentum states, whose main effect is to fill in 
    the   diffraction minimum,  without significantly affecting   the regions 
  left and right to it; consequently,   in these regions, the
 value of  $M({\bf P}, |{\bf p}_{rel}|,\theta_1)$  corresponding to  the arrow  in Fig. \ref{fig10}
 (i.e. to a spectator nucleon at rest),
 can qualitatively be 
 compared with the results shown in Fig. \ref{fig5}  at 
  ${\bf p}_{2z}$=- ${\bf p}_{1z} $= $1.5 fm^{-1}$; in this case one finds
  indeed that the relative momentum of the $np$ pair
  is $|{\bf p}_{rel}|
\simeq 0.75 fm^{-1}$. Thus, in Fig. 5 the region 
where $FSI$ effects 
 are small, correspond to the $2NC$ region  where,  moreover,  the conditions for
the validity of the Schroedinger approach are satisfied.
Note, eventually, that, according to our Glauber 
calculation, as well as to the calculation of 
Ref. \cite{Laget}, the curve shown in Fig. 5 are 
slightly affected by the $pn$ rescattering; thus 
 the dashed line in Fig.5 includes effectively  both the
 $PWA$ and $pp$ rescattering results shown in Fig. 10.
   \item Fig. \ref{fig11}  displays the same as in Fig. \ref{fig10},  but for $\theta_1=90^{o}$.
  The dashed and dot-dashed lines are of course the same as in Fig. \ref{fig9}. In particular,
  at $|{\bf k}_{n}|=1.5 fm^{-1}$, corresponding to the condition $|{\bf k}_{n}|= 2 |{\bf p}_{rel}|$, 
  the $n-pp$ rescattering  is small. Note that in this case, in spite of the fulfillment of the above condition,
  we are not in the {\it two-nucleon correlation} region but rather in the {\it three-nucleon correlation}
  region, for, as shown in the Figure, the momenta of the three-nucleons in the ground-state are of
  comparable size ( $|{\bf k}_{1}| \simeq |{\bf k}_{2}| \simeq |{\bf k}_{n}|/{\sqrt 2}$, $\theta_{12} \simeq 90^{o}$).
 Note that in the $PWA$ the transition form factor both in {\it Process 1} (interaction of $\ga$
  with a proton of a correlated $pp$ pair shown in Figs. 
 \ref{fig4} and \ref{fig5}) and {\it Process 2} (the interaction of $\ga$ with the neutron we are discussing), represent
 the same quantity, namely  the three-body wave function in momentum space (cf. Eqs. \ref{Ti} and \ref{new3}); therefore
  the results presented in
 Figs. \ref{fig10} and \ref{fig11} at   $|{\bf P}|=|{\bf k}_{n}|=0$ have to coincide with the ones
 given in  Fig. \ref{fig3} at the 
 corresponding value of  ${\bf p}_{rel}={\bf k}_{rel}$, as indeed is the case;
  \item Fig. \ref{fig12}  refers to the particular case when the neutron is at rest in the ground-state, so that, after absorbing $\ga$, it leaves the system with
 momentum ${\bf p}_{n}={\bf q}$,  and the two protons are emitted back-to-back in the lab system with 
 $|{\bf p}_1|= |{\bf p}_2|$ and $\theta_1 = 90^{o}$. This is the kinematics also  considered
 in Ref. \cite{GLOCK}. Since, as already pointed out (cf. Section 3.2.1, Eq. (\ref{new3})), in $PWA$
 both {\it Process 1} and {\it Process 2} are described by the same transition form factor, which is nothing but the 
 momentum space three-body wave function, the dashed lines in Figs. \ref{fig4} and \ref{fig12} represent
 the same quantity.
 As a matter of fact, since in  Fig. \ref{fig12}   $|{\bf p}_{rel}|= |{\bf p}_2|$, we see
 that at the highest values of  $|{\bf p}_{rel}|$, the dashed lines in Fig. \ref{fig4} and \ref{fig12}
 are in agreement; note  however that,  due to the effect of the high
 angular momentum states, the dashed line in Fig. \ref{fig12} should not
 exhibit the diffraction minimum seen in Fig. \ref{fig4}.
  It can be seen, as expected from the behaviour of the
 neutron spectral function, that the $pp$ rescattering is very large since the ground-state configuration 
 corresponds to zero neutron momentum and large $pp$ relative momentum.   The fact that the $pp$ rescattering is 
 large,  would not
 represent {\it per se} a serious obstacle in the investigation of the three-body spectral function,
  for, as also stressed in Ref. \cite{GLOCK},  the calculation of the
 $pp$ rescattering is well under control since many years  (see e.g. Ref. \cite{CPS});  unfortunately, also
 the full rescattering effects are very large, with the results that this kinematics is not the optimal one  to
 investigate the three-body wave function. 
The results presented in Fig. \ref{fig12} can qualitatively be compared with the results of Faddeev-like calculations
 of Ref. \cite{GLOCK}, bearing in mind that the latter are restricted to low momentum transfer, i.e. to
  $\sqrt {s} \leq {3M+m_\pi}$,  which means that unlike our case,   for a given value of the three-momentum 
  transfer $|{\bf q}|$,
  there is an upper limit to the  value of $|{\bf p}_{rel}|$.  As far as the  $PWA$ and $pp$ rescattering results
  are concerned, there is en excellent agreement  between our results and the 
$PWIAS$  and $"tG_0"$ results  of Ref. \cite{GLOCK}, respectively,  which is not surprising in view of the similarity
of the wave functions and the treatment of the two-nucleon spectator rescattering adopted in the two calculations;
as for the three-body rescattering
 contribution, there also appear to be a satisfactory agreement between our eikonal-type calculation and the 
 Faddeev results,  provided the values of the three-momentum transfer is large enough; at low values of the
 momentum transfer the eikonal-type approach cannot be applied and a consistent treatment
 of bound and continuum three-nucleon states within the Schr\"odinger approach  is necessary. It should be stressed  that  
  in Ref. \cite{GLOCK} the effect of the three-body rescattering 
 on the process in which $\ga$ is absorbed by a proton at rest and  the proton and the neutron are  emitted back-to-back
  with
   equal
  momenta ${\bf p}_n = -{\bf p}_2 \equiv {\bf p}$,
 has been found to be very small, so that this process
  would be well suited for the investigation of the three-nucleon wave function, being the calculation of the $p-n$
  rescattering well under control, as previously  pointed out;
  it should moreover be emphasized,  that within such a
  kinematics the effect of ${\pi}$ and ${\rho}$ meson exchange  contributions on the transverse transition form factor has also found to be very
  small \cite{GLOCK}.
  However, the {\it s.p.}
   kinematics at $|{\bf p}_m| \geq 2|{\bf p}_{rel}|$
  we have considered seems to be very promising, 
  in view of the smallness of both the spectator-pair and the three-body  rescattering contributions.
 
  \end{enumerate}

\section{Summary and Conclusions}

\begin{figure}[t] 
\epsfxsize 4in \centerline{ \epsfbox{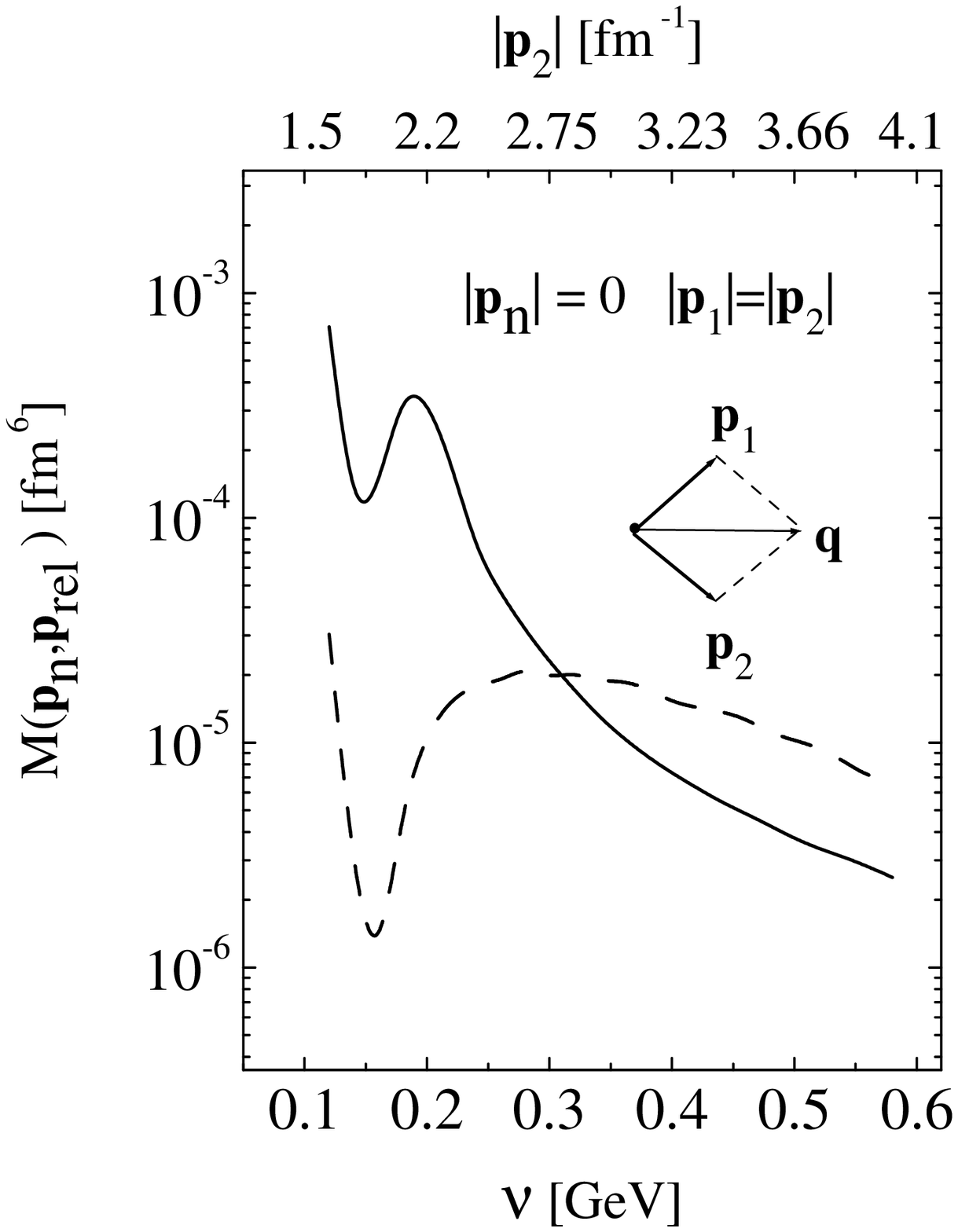}} 
\caption{ The transition form factor $M({\bf 
p_n},{\bf p_{rel}},{\bf q})$ (Eq.\ref{oneH}) 
calculated in the {\it Symmetric kinematics} 
(\cite{Laget}): ${\bf p}_1+{\bf p}_2$ = ${\bf 
q}$, ${\bf p}_n=0$, $|{\bf p}_1|$= $|{\bf 
p}_2|$=$\sqrt 
{\frac{1}{4}(\nu+M_3-M_n)^2-M_p^2}$, ${\bf 
p}_{rel}= {\bf q}/2-{\bf p}_2$, $\theta_{12}= 2 
\arccos {\frac {|{\bf q}|} {2|{\bf p}_2|}}$. 
 The dashed line corresponds to the Plane Wave Approximation ($PWA$) (plane waves for the three-nucleons,
  $Process \, (a)$ in Fig. \ref{fig2}), whereas 
the full line includes the $pp$ rescattering 
($Process\, (a)$ + $Process\, (b)$ in Fig. 
\ref{fig2}). The value of $|{\bf q}|$ corresponds 
to $\epsilon_e = 2 GeV$ and $\theta_e=15^{o}$ 
(cf. Eq. \ref{qu}), and the range of its 
variation with $\nu$ is 
 $0.52\ GeV/c \le |{\bf q}|\, \le 0.75\ GeV/c$. 
 Three-body wave function from Ref. \cite{rosati};  $AV18$ interaction  \cite{AV14}.}
\label{fig4} 
\end{figure}

\begin{figure}[t] 
\epsfxsize 4in \centerline{ \epsfbox{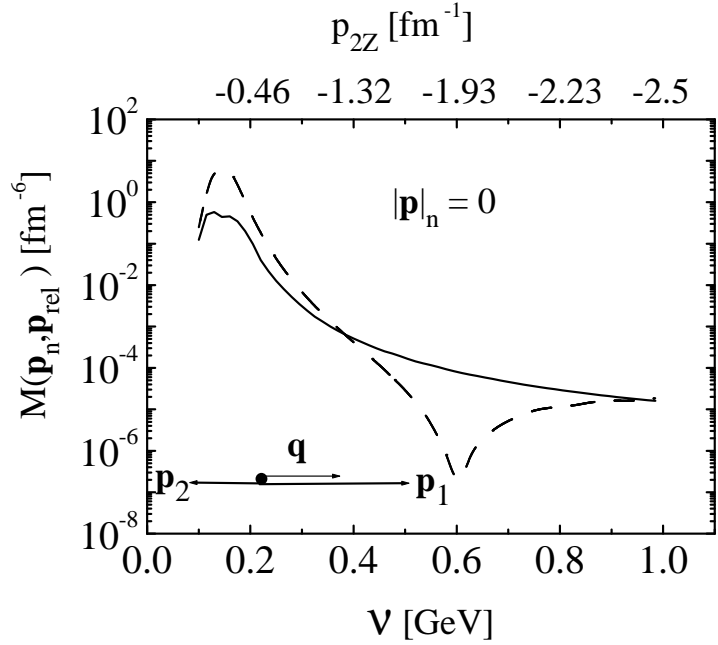}} 
\caption{ The transition form factor $M({\bf 
p_n},{\bf p_{rel}},{\bf q})$ (Eq.\ref{oneH}) 
calculated in the {\it Super-parallel kinematics} 
(${\bf p}_{1\perp}={\bf p}_{2\perp}$=0, 
${p}_{1z}+{ p}_{2z}= |{\bf q}|$, with ${\bf q} 
\parallel z$). 
 The dashed line corresponds to the Plane Wave Approximation ($PWA$) (plane waves for the three-nucleons,
  $Process \, (a)$ in Fig. \ref{fig2}), whereas 
the full line includes the $pp$ rescattering 
($Process\, (a)$ + $Process\, (b)$ in Fig. 
\ref{fig2}). The value of $|{\bf q}|$ corresponds 
to $\epsilon_e = 2 GeV$ and $\theta_e=15^{o}$ 
(cf. Eq. \ref{qu}) and 
 the range of its  variation with $\nu$  is
 $0.52\ GeV/c \le |{\bf q}|\, \le 1.0\ GeV/c$. Three-body wave function from Ref. 
\cite{rosati}; $AV18$ interaction 
  \cite{AV14}.}
\label{fig5} 
\end{figure}

\begin{figure}[h] 
\begin{center}
    \includegraphics[height=0.50\textheight]{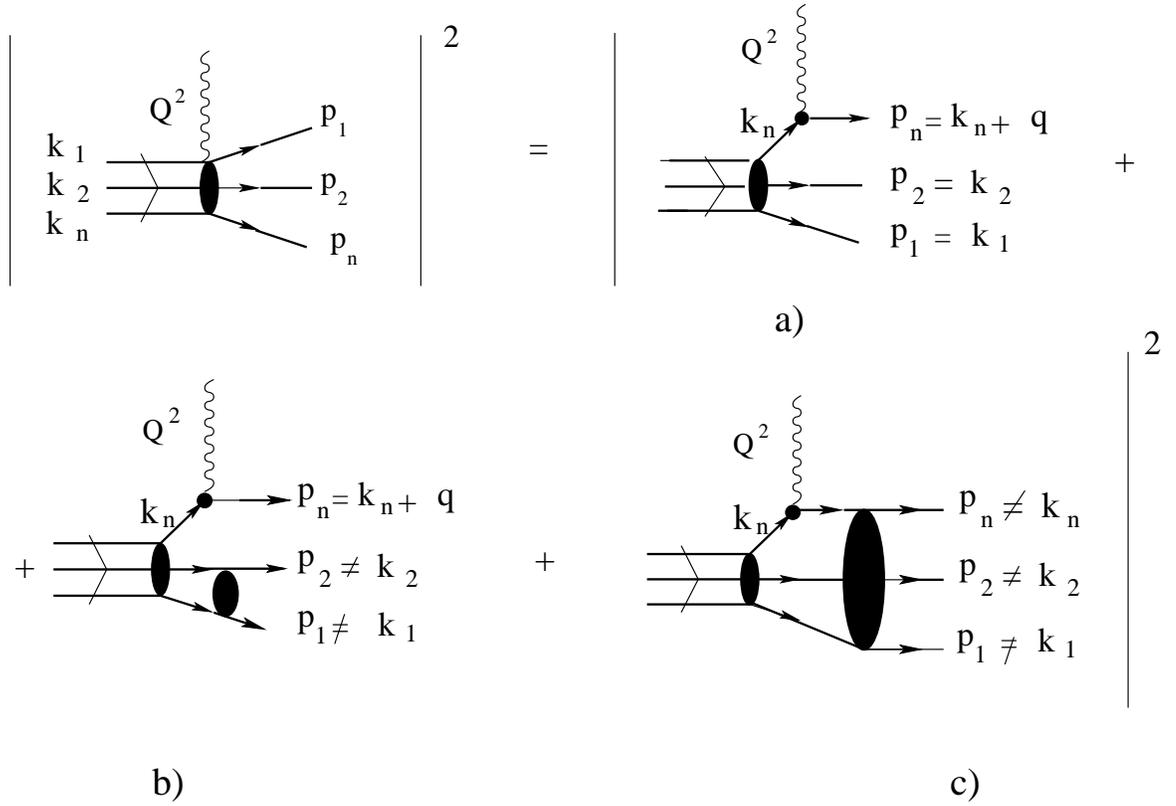}
\vskip -0.25cm \caption{Schematic representation 
of the various processes contributing to the 
 reaction  $\pp$ when $\gamma^{*}$ is absorbed by the neutron, and the two protons are 
 emitted in the continuum: (a) denotes the Plane Wave Approximation ({\it PWA}), (b) the 
 {\it  pp  rescattering}, (c) the {\it three-body rescattering}.
The sum of contributions $a)$ and $b)$ is 
referred to by some authors as the 
 {\it Plane Wave
 Impulse Approximation (PWIA)}; in Ref. \cite{GLOCK} $PWIA$  is used, on the contrary, 
  to denote our (symmetrized)  $PWA$
  approximation.
 In the rest of this paper we shall be using the term $PWA$ to denote process $a)$, and the term  {\it pp rescattering}
  to denote process $b)$. Note moreover, that in Ref. \cite{GLOCK}
our $pp$ rescattering contribution is called 
 $"tG_0"$.}
\label{fig6} 
\end{center}
\end{figure}

\begin{figure}[t] 
\epsfxsize 4in \centerline{ \epsfbox{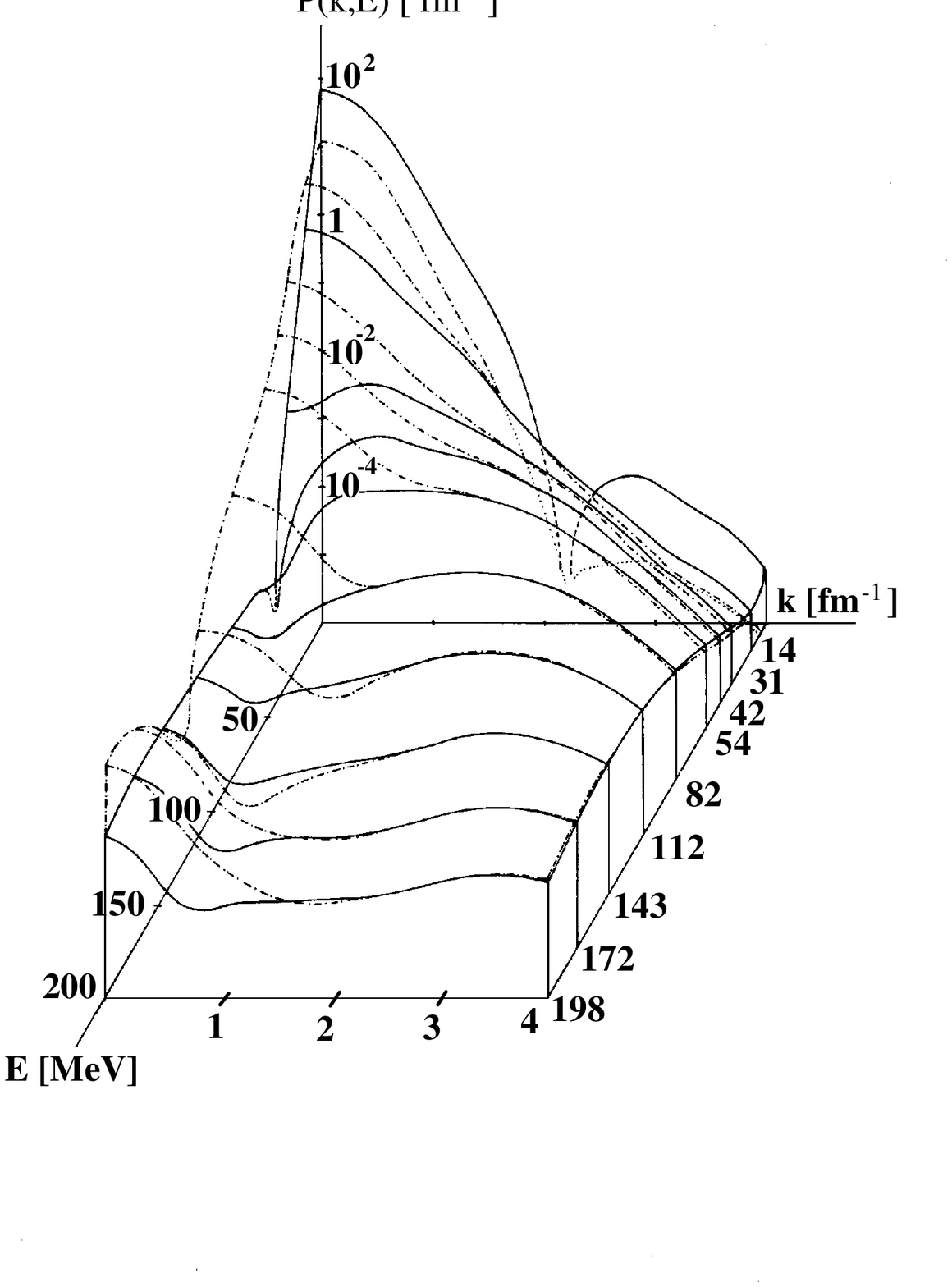}} 
\caption{The neutron Spectral Function in $^3He$ 
($k\equiv |{\bf k}_n|$) corresponding to the 
three-body channel $^3He \rightarrow (pp)-n$ . 
The dot-dashed line represents the $PWA$, whereas 
the full line includes the 
 proton-proton  rescattering. Three-nucleon
 wave function from \cite{CPS}; Reid Soft Core interaction
\cite{Reid} (adapted from. Ref. \cite{CPS}).} 
\label{fig7} 
\end{figure}

In this paper we have investigated the effects of 
the Final State Interaction (FSI) in the process 
$^3He(e.e'2p)n$ 
 using
 realistic  three-nucleon wave functions  which, being the exact solution of the
 Schr\"odinger equation, incorporate all types of correlations
 generated by modern NN potentials.
 We have taken into account FSI effects, treating the three-nucleon rescattering within an improved eikonal approximation,
 which allows one to consider the two-nucleon emission processes at high momentum transfer, 
 also when  $\sqrt {s} \ge {3M +m_\pi}$, i.e. above the kinematical boundary imposed by Faddeev-like calculations.
We reiterate once again that  our aim was restricted to the development of a theoretical approach for the
treatment of FSI effects in two-nucleon emission processes off the three-body system, and to the
investigation of the effects produced by FSI  in various kinematical regions. We did not discuss,
other final state effects, e.g. MEC, which clearly have to be taken into account when theoretical predictions are
compared with   experimental data. 
We have been guided by the idea that if a kinematical region could be found, where the effects of FSI are minimized,
this would represent a crucial advance  towards the investigation of both GSC and current operators.
Basically we have considered two different mechanisms leading 
 to the two-proton emission process:
 \begin{enumerate}
 
 \item{\it Mechanism 1}, \, in which $\ga$ is absorbed by a correlated $pp$ pair. This
   mechanism, which  is  
 the one usually  considered in the case of complex nuclei,  has  been 
 previously analyzed  in Ref. 
 \cite{Laget} within  a particular kinematics, the so called
  {\it symmetric kinematics}, according to which
   ${\bf p}_{1}+{\bf p}_2$=
  ${\bf q}$,  ${\bf p}_n=0$. In $PWA$, such a kinematics selects  the  ground-state wave function configuration
  in which 
   ${\bf k}_{1} \simeq  -{\bf k}_2$, 
    ${\bf k}_n=0$ (the {\it two-nucleon correlation} ($2NC$) configuration).
     Our calculations confirm the results of Ref. \cite{Laget}, namely that
      the FSI due to the  $n-(pp)$ rescattering is 
     very small, whereas  the $pp$ rescattering is extremely large, 
      and fully distorts the direct link between the ground-state wave function and 
      the cross section, which holds  in $PWA$ (cf. Fig. \ref{fig4}).
      Thus,  the {\it symmetric
    kinematic}  does not appear extremely useful to investigate the three-nucleon wave function.
     A more interesting kinematics is the  
     {\it super-parallel kinematics}
(${\bf p}_{1\perp}={\bf p}_{2\perp}=0 ,{p}_{1z}+{ 
p}_{2z}= |{\bf q}|, {\bf q} \parallel z$) which, 
as demonstrated in Fig. \ref{fig5}, shows that, 
particularly at high values of the momenta of the 
detected protons, the effects from the FSI are 
strongly reduced; 

\item {\it Mechanism 2},\, in which $\ga$ is absorbed by the neutron,
 and the two protons are detected. We have shown that if the $pp$  rescattering  
 is taken into account and the one between the 
neutron and the protons disregarded, the cross 
section 
 depends 
upon the {\it Vector Spectral Function} 
${\mathcal P}_1({\bf k}_n, {\bf p}_{rel})= 
\frac{M_N|{\bf p}_{rel}|}{2} M^{pp}({\bf k}_n, 
{\bf p}_{rel})$. This Spectral Function, 
 unlike the usual one (see e.g Ref. \cite{CPS}),
 depends not only upon $|{\bf k}_n|$ and   $|{\bf p}_{rel}|=\sqrt{2ME^*}$, but also upon
  the angle $\theta$ between them. 
 By analyzing the behaviour of   ${\mathcal P}_1({\bf k}_n, {\bf p}_{rel})$
  (cf. Fig. \ref{fig9}) we
 have  demonstrated that: i) when  the relation $|{\bf k}_{n}|\simeq  2|{\bf p}_{rel}|$ holds, 
the FSI due to $pp$ rescattering is very small, 
and ii) the dominant configuration in the 
ground-state wave function is the $2NC$ one, in 
which 
   ${\bf k}_{1}\simeq -{\bf k}_2$, 
    ${\bf k}_n=0$ (Eq. \ref{5teen1}); we have also found that when $\theta \simeq 0^{o} (180^{o})$,
     the smallness
     of $pp$ rescattering actually extends to a wide region 
     characterized by 
     $|{\bf k}_{n}|> 2|{\bf p}_{rel}|$,
 where the  $2NC$  configuration is still the dominant
 one (cf. Appendix). Such a picture is not in principle withstanding when
    the three-nucleon $n-(pp)$ rescattering is taken into account, since in this case
    the concept of neutron momentum before interaction ${\bf k}_n$ 
    has to be abandoned in favor of  the concept of {\it missing momentum} 
      ${\bf p}_{m}$ = ${\bf p}_{n} - {\bf q}$ = -(${\bf p}_{1}+{\bf p}_{2}$), which
    equals ${\bf k}_n$ only when  the three-body rescattering is disregarded. 
    However, our consideration of  the three-body rescattering, 
     clearly show that in the case of the 
      {\it super-parallel kinematics}, 
      both the three-body and two-body rescattering are negligible
       when  $|{\bf p}_{m}|\ge 2|{\bf p}_{rel}|$,
      which means that in this region 
       $|{\bf p}_{m}|\simeq |{\bf k}_{n}|$.
        The three-body rescattering is on the contrary
      very relevant when  $|{\bf p}_{m}| < 2|{\bf p}_{rel}|$, 
      both in the case of the {\it super-parallel kinematics},  and particularly 
      when the two protons are detected with their relative momentum perpendicular
       to the direction of ${\bf q}$ 
      (cf. Fig. \ref{fig11}). Within {\it Mechanism 2} we have, as in Ref. \cite{GLOCK},
       also considered the process  in which $\ga$ is absorbed by a neutron at rest
     and the two protons are emitted back-to-back in the direction perpendicular to the direction of 
     the momentum transfer  ${\bf q}$, which means ${\bf p}_m =0$. In  this case,
      the  effects
     from the FSI appear to be very different from the ones considered in the two previous cases,
     namely, unlike the {\it symmetric kinematics}  (cf.  Figs. \ref{fig4} and \ref{fig5}),
     where only the FSI between the two active protons played a substantial role, here {\it both} the 
     $pp$ rescattering and the three-body  $n-(pp)$ rescattering are very large (cf. Fig. \ref{fig12}).
     \end{enumerate}
     
     We can summarize the main results we have obtained in the following way:
     
     \noindent i)
      if $\ga$ is absorbed by a correlated proton pair,  with the spectator neutron  at rest and  the two protons 
      detected with their relative momentum perpendicular to the direction of ${\bf q}$ ({\it symmetric kinematics}), 
     the leading FSI is the $pp$
     rescattering, with  the $n-(pp)$ rescattering playing only a minor role. 
     In such a case, however,  the  $pp$ rescattering fully destroys 
     the direct link between the ground-state wave function and  the cross section, occurring in the $PWA$;
      if the protons, on the contrary, are   detected 
     with their relative momentum parallel to ${\bf q}$ ({\it super-parallel kinematics}),
      the effects of the $pp$ rescattering 
     is appreciably suppressed, particularly at high values of the momenta of the detected protons;
     
     \noindent ii) if $\ga$ is absorbed by an uncorrelated neutron at rest and the two correlated protons are emitted
     back-to-back with ${\bf p}_1$ = - ${\bf p}_2$,  ${\bf p}_{rel}= {\bf p}_1 \perp {\bf q}$, {\it  both}
      the $pp$ 
     and the  $n-(pp)$ rescattering are very large;
     
     \noindent iii) if $\ga$ is absorbed by a neutron and the two protons are detected in the
      {\it super-parallel kinematics}, one has the following situation:  if $|{\bf p}_m| <2 |{\bf p}_{rel}|$ {\it both} the $pp$ and the  $n-(pp)$
     rescattering are large; if, on the contrary,  $|{\bf p}_m| \ge 2 |{\bf p}_{rel}|$ they are both small and the cross section
     can be directly linked to the three-body wave function in momentum space.

In conclusion it appears that the {\it 
super-parallel kinematics} with 
 $|{\bf p}_m| \ge 2 |{\bf p}_{rel}|$ could represent a powerful tool to investigate the structure of
 the three-body wave function in momentum space. Experimental data in this kinematical region would be therefore highly desirable.

{\bf Acknowledgments} 

This work was partially supported by the 
Ministero dell'Istruzione, Universit\`{a} e 
Ricerca (MIUR), 
 through the funds COFIN01.
We are indebted to A. Kievsky for making 
available the variational three-body wave 
functions of the Pisa Group and to W. Gl\"ockle 
for useful discussions and for supplying the 
results of the $^3He(e,e'pp)n$ calculations prior 
to publications. We thank L. Weinstein and B. 
Zhang for the information they provided on the 
CLAS preliminary results on two-proton emission 
off $^3He$. L.P.K. is indebted to the University 
of Perugia and INFN, Sezione di Perugia, for warm 
hospitality and financial support.

\newpage

\centerline {\bf APPENDIX}

\centerline{\bf THE BASIC CONFIGURATIONS IN THE 
VECTOR SPECTRAL FUNCTION}

Let us investigate the basic configuration in the 
Vector Spectral Function. To this end, we will 
first consider the three-body wave function in 
momentum space or, in other words, the {\it 
Vector Spectral Function} 
 in the Plane Wave Approximation,
when the momenta of the three-nucleons satisfy 
the relation 
\be
{\bf k}_1 + {\bf k}_2 + {\bf k}_n = 0 \label{B1} 
\ee i.e. 
\be
{\bf k}_{1(2)}=-\frac12 {\bf k}_{n} \pm {\bf 
k}_{rel} 
 \label{B2}
\ee 

\noindent where ${\bf k}_{rel} = \frac{{\bf 
k}_1-{\bf k}_2}{2}$ is the relative momentum. One 
thus have 
\be
|{\bf k}_{1(2)}|= \sqrt{{\bf 
k}_{rel}^2+\frac{{\bf k}_n^2}{4} \mp |{\bf 
k}_{rel}||{\bf k}_n| cos\theta} \label{B4} \ee 

The above relation illustrates that: 
\begin{enumerate}
\item if  $ |{\bf k}_{n}| = 2|{\bf k}_{rel}|$,  $|{\bf k}_{1}|$ =0, ${\bf k}_{n}= -{\bf k}_{2}$, 
when $\theta =0^o$, and 
 $|{\bf k}_{2}|$ =0, ${\bf k}_{n}= -{\bf k}_{1}$, when $\theta =180^o$, which
 represents the {\it two-nucleon correlation} ($2NC$) configuration. It should  be pointed out that the scalar  condition
 $ |{\bf k}_{n}| = 2|{\bf k}_{rel}|$ alone does not suffice  to uniquely specify the ground-state
 configuration, in particular the $2NC$ configuration. As a matter of fact,
 when  $ |{\bf k}_{n}| = 2|{\bf k}_{rel}|$  but $\theta=90^o$, one has 
 $|{\bf k}_1| = |{\bf k}_2|=\sqrt{2}|{\bf k}_{rel}|$ which represents
 a typical {\it three-nucleon correlation} ($3NC$) configuration, for 
 all of the three-nucleons have comparable and high momenta. However,  it can be seen from
  Fig.\ref{fig9} that   such a three-nucleon configuration is strongly suppressed, and 
  $P(k_n,E)$ is  mainly governed by the $2NC$  configuration.
  \item If  $ |{\bf k}_{n}| \neq 2|{\bf k}_{rel}|$, we do not stay in the $2NC$ configuration, but it can easily
  be checked that if  $|{\bf k}_{n}| > 2|{\bf k}_{rel}|$ and $\theta \simeq 0^o (180^0)$,  $|{\bf k}_n|$ is still comparable
  with  $|{\bf k}_{2(1)}|>>|{\bf k}_{1(2)}|$.
  \end{enumerate}
  
  The above picture is distorted by the $pp$ rescattering, but, as demonstrated in Fig. \ref{fig9},  only for: i)  
   $|{\bf k}_{n}| < 2|{\bf k}_{rel}|$ and ii)
    $|{\bf k}_{n}| > 2|{\bf k}_{rel}|$, $\theta \simeq 90^o$.

\newpage

\pagestyle{empty} 


\newpage


 \begin{figure}[h] 
 \vskip .7cm
\begin{center}
    \includegraphics[height=0.70\textheight]{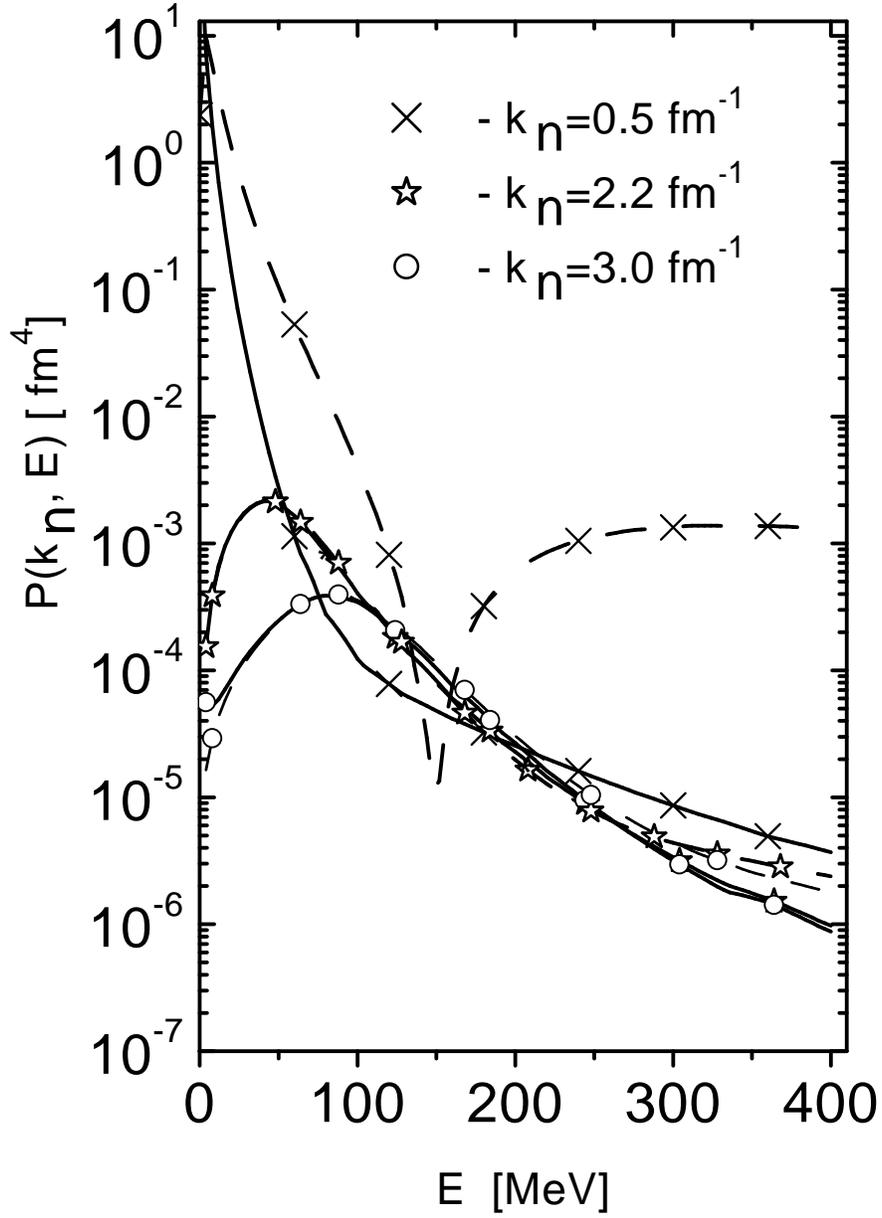}
\caption{The same as in Fig. \ref{fig7} but with 
the three-body wave function from Ref. 
\cite{rosati} and the $AV18$ interaction 
\cite{AV14} . The E-dependence of the Spectral 
Function is shown for three values of the neutron 
momentum. The dashed line represents the $PWA$ 
results, whereas the full line includes the $pp$ 
rescattering.} \label{fig8} 
\end{center}
\end{figure}

 \begin{figure}[h] 
 \vskip -.1cm
\begin{center}
\includegraphics[height=0.35\textheight]{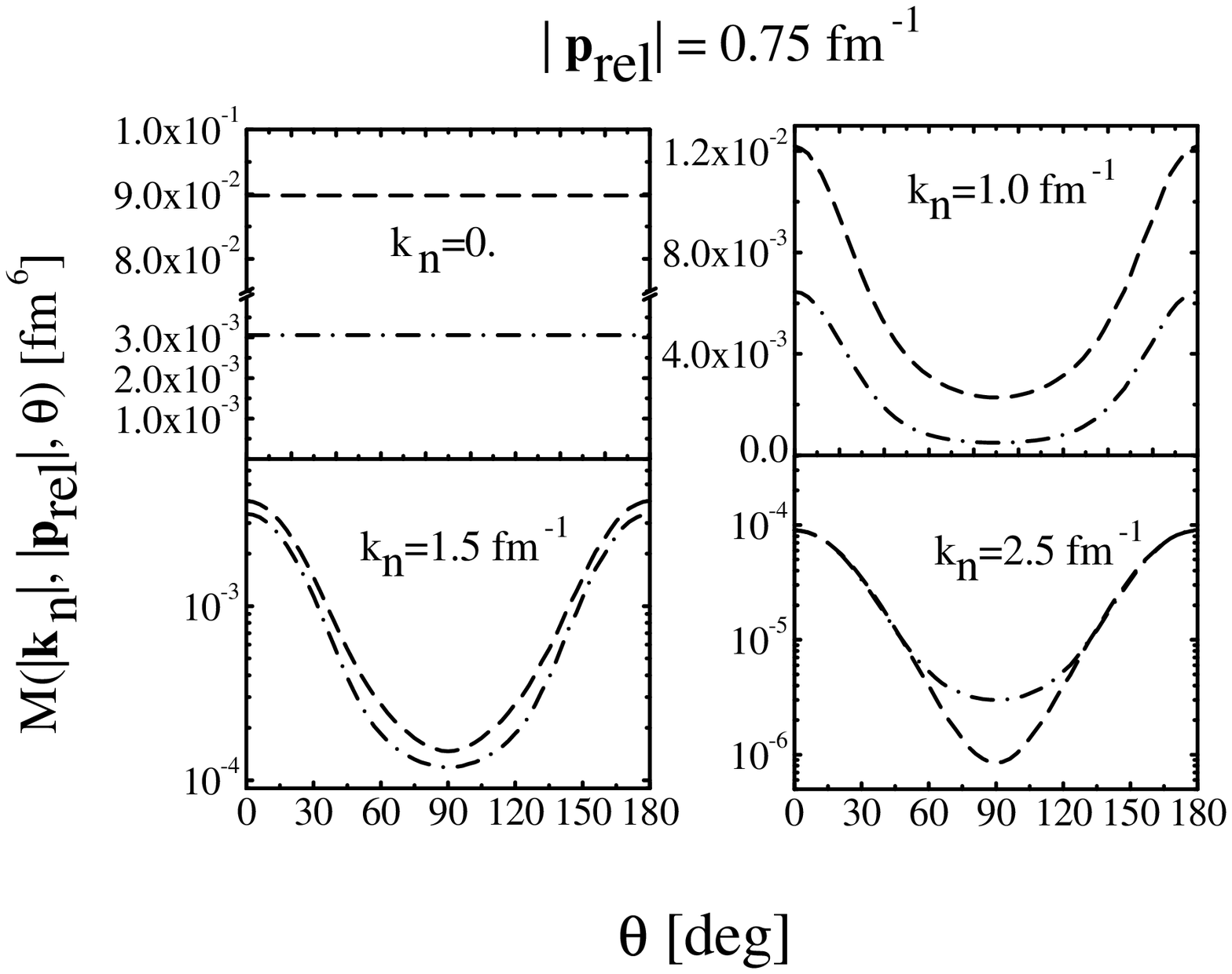}
\end{center}
\begin{center}
\includegraphics[height=0.35\textheight]{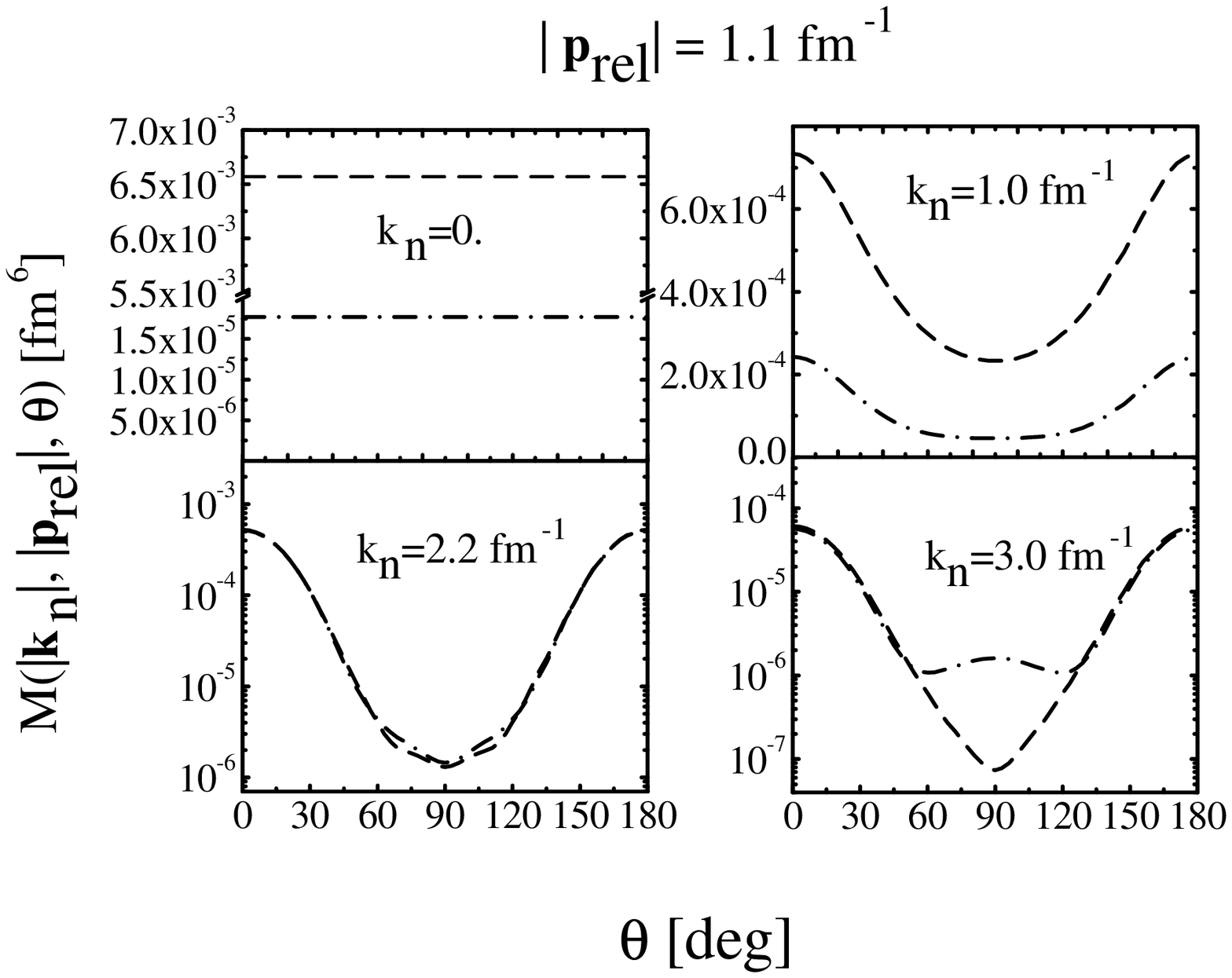}
\caption{{\it Top panel}. Dashed line: the 
transition form factor 
 $ M^{pp}({\bf k}_n, {\bf p}_{rel}) \equiv\frac{2}{M_N|{\bf p}_{rel}|}
 {\mathcal P}_1({\bf k}_n, {\bf p}_{rel})$ (see Eq. \ref{5teen1})
 plotted {\it vs} the angle  ${\theta}$  between ${\bf k}_n$ and ${\bf p}_{rel}$
for various values of ${k}_{n} \equiv |{\bf 
k}_n|$ and $|{\bf p}_{rel}|=0.75 fm^{-1}$ (the 
corresponding "excitation energy" of the $pp$ 
pair is $E^{*}=|{\bf p}_{rel}|^2/M_N \simeq 23 
MeV $). Dot-dashed line: the same quantity in the 
$PWA$, i.e. disregarding the $pp$ rescattering. 
Three-body wave function from Ref. \cite{rosati}; 
$AV18$ interaction \cite{AV14}.\newline 
\hspace*{1cm} {\it Bottom panel}: the same as in 
{\it Top panel} for 
 $|{\bf p}_{rel}|=1.1 fm^{-1}$  ($E^{*}=|{\bf p}_{rel}|^2/M_N \simeq 50 MeV $).}
\label{fig9} 
\end{center}
\end{figure}


      
\begin{figure}[h]
 \vskip -3cm
  \begin{center}
    \includegraphics[height=0.45\textheight]{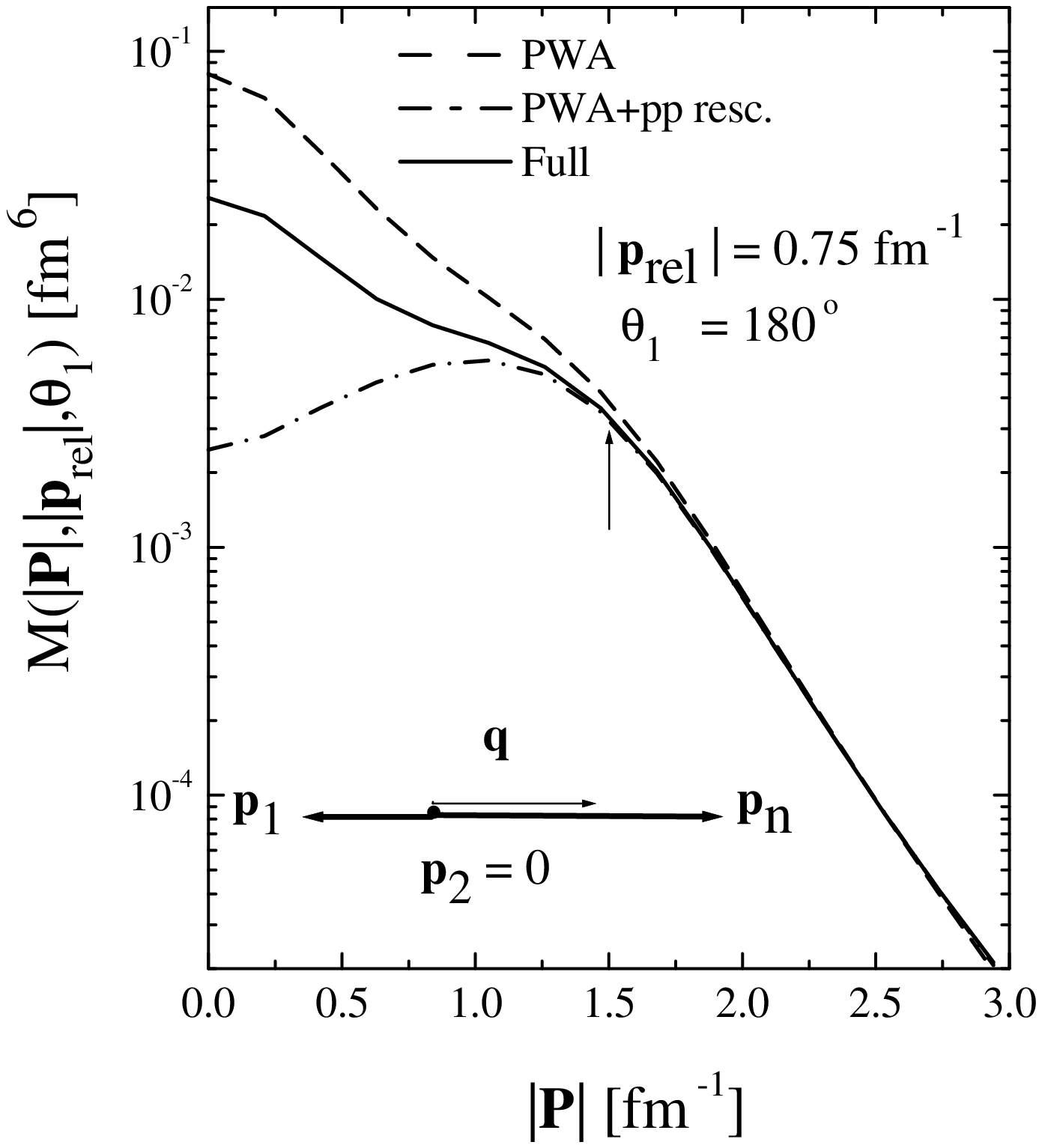}
    \caption{The transition form factor  
    $M(|{\bf P}|,|{\bf p}_{rel}|,\theta_1)$ = $M({\bf p_n},{\bf p_{rel}},{\bf q})$
     (Eq.\ref{oneH}),
 plotted {\it vs} the  two-nucleon Center-of-Mass (or {\it missing}) 
  momentum $|{\bf P}|$ = $|{\bf p}_1 + {\bf p}_2|$ = $|{\bf p}_m|$
= $|{\bf q} - {\bf p}_n|$ for fixed values of the relative momentum $|{\bf p}_{rel}|$ and  the angle $\theta_1$ 
between ${\bf q}$ 
     and ${\bf p}_{rel}$.  
   The dashed and  dot-dashed lines represent, as in Fig. \ref{fig9},   the
   $PWA$ (Eq. \ref{Ti}) and the $PWA$ plus  $pp$ rescattering 
   (Eq.\ref{oneH}), respectively, whereas  the full
    line includes also the $n-(pp)$
   rescattering according to Eq. \ref{MG}.
     The arrow and the momentum 
    vector balance, which refer to the dashed and the dot-dashed lines,  
   denote
    the point  where Eq. \ref{9teen} is satisfied,  i.e.  $|{\bf k}_{n}|$=2$|{\bf p}_{rel}|$=$1.5 fm^{-1}$,
   ${\bf k}_1 \simeq -{\bf k}_{n}$, ${\bf k}_2 \simeq 0$; when  $\theta_1 =0^{o}$, the behaviour of 
    the dashed and dot-dashed lines is  exactly the same, with the arrow denoting
    in this case  
    the configuration with  $|{\bf k}_{n}|$=2$
    |{\bf p}_{rel}|$,  ${\bf k}_2 \simeq -{\bf k}_{n}$, ${\bf k}_1 \simeq 0$. 
    The inclusion of the $n-(pp)$
    rescattering destroys in principle  the  $\theta_1 =0^{o}-180^{o}$ symmetry, but,
     as explained in the text, the asymmetry
    generated by our  calculation is  very mild.
     For  $|{\bf P}| > 1.5 fm^{-1}$, the ground-state momentum balance is always similar 
      to the $2NC$ configuration ($|{\bf k}_n| \simeq |{\bf k}_1|$,  $|{\bf k}_2| \ll |{\bf k}_1|$),   
     whereas  for  $|{\bf P}| < 1.5 fm^{-1}$,  the  configuration is far from the $2NC$ one.           
     Three-nucleon
 wave function from \cite{rosati};
  AV18 interaction \cite{AV14}.}
      \label{fig10}
    \end{center}
\end{figure}


         \begin{figure}[h] 
 \vskip 1cm
  \begin{center}
    \includegraphics[height=0.50\textheight]{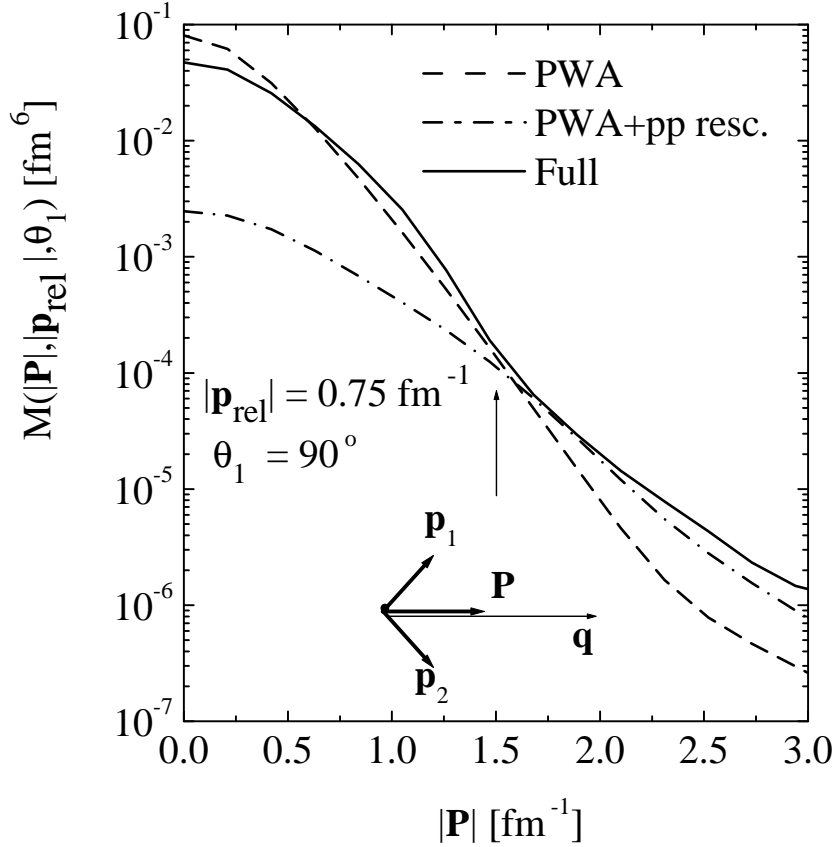}
    \caption{The same as in Fig. 
    \ref{fig10} but for   $\theta_1 =90^{o}$ (cf. Fig. \ref{fig9}, 
    as far as the dashed and dot-dashed
    lines are concerned).
     The arrow and the momentum vector balance  correspond to the 
     point 
     where Eq. \ref{9teen} is satisfied  ($|{\bf k}_{n}|$=2$|{\bf p}_{rel}|$=$1.5 fm^{-1}$), but since   $\theta_1 \neq 
    0^{o}$ ($180^{o}$), we do not stay now in the  {\it 
    two nucleon correlation} region, but rather in the  {\it 
    three- nucleon correlation} region, for 
     $|{\bf k}_2| \simeq |{\bf k}_{1}| \simeq |{\bf k}_n|/{\sqrt 2} \simeq 1.1\ fm^{-1}$ (cf. Appendix ).
      Three-nucleon
 wave function from \cite{rosati}; AV18 interaction \cite{AV14}.}
     \label{fig11}
    \end{center}
\end{figure}                                  
                                          
  \begin{figure}[h]                                      
 \vskip 0.1 cm
  \begin{center}
    \includegraphics[height=0.50\textheight]{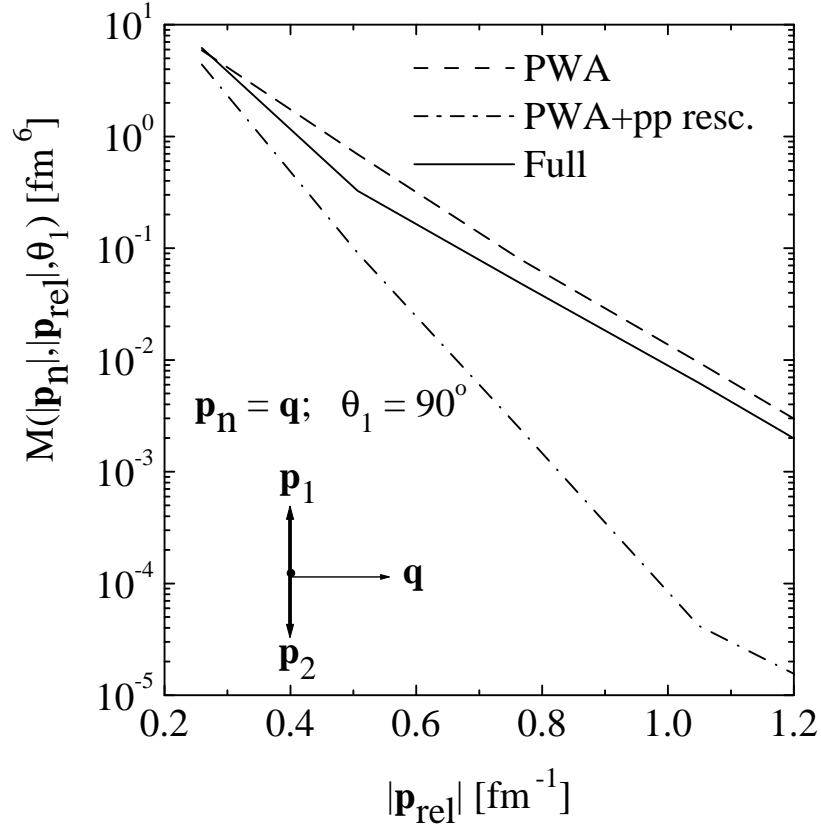}
    \caption{
     The transition form factor  $M(|{\bf P}|,|{\bf p_{rel}}|,\theta_1)$= $M({\bf p_n},{\bf p_{rel}},{\bf q})$
     (Eq.\ref{oneH}), plotted {\it vs}  $|{\bf p}_{rel}|$ for  ${\bf p_{n}}={\bf q}$,  $\theta_1=90^{o}$.
     The process  corresponds  to the absorption
     of $\gamma^*$ by a neutron at rest followed by the emission of  two protons with momenta  ${\bf p}_1 = -{\bf p}_2$
     ({\it back-to-back} protons).
     Three-nucleon 
 wave function from \cite{rosati}; AV18 interaction.}
      \label{fig12}
    \end{center}
\end{figure}
 \end{document}